\documentclass{elsart}

\usepackage{cite,graphicx,amsmath,amsfonts,amssymb,mathrsfs}

\newcommand{\ie}{{\it i.e.}}

\newcommand{\eg}{{\it e.g.}}

\newcommand{\cf}{{\it cf.}}

\newcommand{\eq}{Eq.}
\newcommand{\eqs}{Eqs.}
\newcommand{\Def}{Def.}
\newcommand{\fig}{Fig.}

\newcommand{\property}{Property}
\newcommand{\Ref}{Ref.}
\newcommand{\Sec}{Sec.}
\newcommand{\Subsec}{Subsec.}

\newtheorem{definition}{Definition}
\newtheorem{Prop}{Property}

\begin{document}

\begin{frontmatter}

\begin{flushright}
{\small TUM-HEP-408/01}
\end{flushright}

\title{A combined treatment of neutrino decay and neutrino oscillations}

\author[labela]{Manfred Lindner\thanksref{label0}},
\thanks[label0]{E-mail: lindner@physik.tu-muenchen.de}
\author[labela,labelb]{Tommy Ohlsson\thanksref{label1}},
\thanks[label1]{E-mail: tohlsson@physik.tu-muenchen.de, tommy@theophys.kth.se}
\author[labela]{Walter Winter\thanksref{label2}}
\thanks[label2]{E-mail: wwinter@physik.tu-muenchen.de}

\address[labela]{Institut f{\"u}r Theoretische Physik, Physik
Department, Technische Universit{\"a}t M{\"u}nchen,
James-Franck-Stra{\ss}e, D-85748 Garching bei M{\"u}nchen, Germany}

\address[labelb]{Division of Mathematical Physics, Theoretical
Physics, Department of Physics, Royal Institute of Technology (KTH), SE-100
44 Stockholm, Sweden}

\date{\today}
         
\begin{abstract}
Neutrino decay in vacuum has often been considered as an alternative
to neutrino oscillations. Because non-zero neutrino masses imply the
possibility of both neutrino decay and neutrino oscillations, we present a
model-independent formal treatment of these combined scenarios. 
For that, we show for the example of Majoron decay that in many cases decay
products are observable and may even oscillate.
Furthermore, we construct a minimal scenario in which we study the
physical implications of neutrino oscillations with intermediate decays. 
\end{abstract}

\end{frontmatter}

\section{Introduction}

The most favorable alternative for fast neutrino decay is to
introduce an effective decay Lagrangian which couples the neutrino fields
to a massless boson carrying lepton number, \ie, a complex
scalar field or a Majoron field
\cite{Zatsepin:1978iy,Chikashige:1980qk,Gelmini:1981re,Pakvasa:1999ta}. One
possibility for such a Lagrangian for the case of Majoron decay is\footnote{We
are not interested in $\nu_i \rightarrow \nu_i$ transitions described by
$g_{ii} \neq 0$, since they are forbidden for kinematical reasons. Furthermore,
because the $g_{ij}$'s are usually assumed to be small, only the first order
processes in the Lagrangian $\mathcal{L}_{\rm int}$ will be relevant to our
discussion.}
\begin{equation} \label{int2} \mathcal{L}_{\rm int} = \underset{i
\neq j}{\sum_{i} \sum_{j}} g_{ij} \overline{\nu_{j,L}^{c}} \nu_{i,L} J,
\end{equation}
where the $\nu$'s are Majorana mass eigenfields and $J$ is a Majoron field. 

Neutrino decay has been studied as an alternative to neutrino
oscillations, especially for atmospheric
\cite{Acker:1992eh,Barger:1998xk,Barger:1999bg,Fogli:1999qt} and solar
\cite{Acker:1992eh,Choubey:2000an,Bandyopadhyay:2001ct} neutrinos. So far,
either decay only or some sequential combination of decay and oscillations has
been considered (\eg, MSW-mediated solar neutrino decay
\cite{Bandyopadhyay:2001ct}). 
The values of individual $g_{ij}$'s or combinations were restricted, for
instance, from double beta decay, pion and kaon decays,
or supernovae
\cite{Barger:1982vd,Acker:1992ej,Kachelriess:2000qc,Tomas:2001}. For pure
neutrino decay without oscillations the constraints on the coupling constants
have been used to derive conditions for other parameters. For example,
for atmospheric neutrinos $\Delta m^2 \ge 0.73 \, {\rm eV}^2$ was found in \Ref
\cite{Barger:1998xk}. Since this is incompatible with the known (or usually
assumed) mass hierarchy of active neutrinos, decay into a sterile neutrino was
suggested.
However, massive neutrinos allow neutrino decay as well as neutrino
oscillations. In this paper, we present a combined
model-independent treatment of neutrino decay and neutrino oscillations. We
refer to two different cases:
\begin{description}
\item[Invisible decay:]
Decay into neutrinos or other particles which are not observable (\eg, sterile
decoupled neutrinos\footnote{In this paper, {\em sterile
decoupled neutrinos} refers to neutrinos which can be described by a block
diagonal mixing matrix such as
\begin{equation}
 \label{decoupled}
 U= \left( \begin{array}{cc} U_{3 \times 3} & 0 \\ 0 &
U_{n_S \times n_S} \end{array} \right),
\end{equation}
where $n_S$ is the number of
sterile neutrinos, $U_{3 \times 3}$ the active neutrino mixing matrix, and
$U_{n_S \times n_S}$ the sterile neutrino mixing matrix.}).
\item[Visible decay:]
Decay into neutrinos which are, in principle, observable (\eg, active
neutrinos). 
\end{description}

Our paper is organized as follows: In \Sec~\ref{sec:majoron}, we will
introduce a Majoron decay model for illustration, in order to show two
important properties: First, as already mentioned before (\eg, in
\Ref \cite{Acker:1992eh}), we may be able to detect active decay
products. Second, we will show that in such a model, decay
products can, in principle, oscillate.
In \Sec~\ref{sec:oap}, we will develop a model-independent
operator framework for the calculation of transition probabilities and
describe its properties. Furthermore, we will show that the phase
relationships given by the $S$-matrix approach of quantum field theory are
satisfied.
The transition probability for {\em invisible decay} as a generalization of the
neutrino oscillation formula will be derived in
\Sec~\ref{sec:invis}. As an example, we will apply this to atmospheric
neutrinos.
In the following section, \Sec~\ref{sec:vis}, this formalism will be
used for {\em visible decay} in the approximation of almost stable decay
products. Again, we will make an application to atmospheric neutrinos.
In \Sec~\ref{sec:nodp}, we will, for illustration, construct a minimal model in
which decay products oscillate.
Finally, in \Sec~\ref{sec:sc}, we will present a summary as well as
our conclusions.

\section{Majoron models}
\label{sec:majoron}

Neutrino decay through emission of a massless Nambu--Goldstone boson is
currently among the most favorable alternatives for neutrino decay scenarios.
Therefore, we will use it as an introductory example for our framework.

\subsection{Dynamics of Majoron decay}

For quantitative estimates we will investigate decay by Majoron emission
described by an interaction Lagrangian with Yukawa (scalar or pseudoscalar)
couplings (\cf, \Ref \cite{Kim:1990km}),
\begin{equation}
 \mathcal{L}_{\rm int} = {g_1 \over 2} \bar{\nu}_1 \nu_2 J + {g_2 \over 2}
 \bar{\nu}_1 i \gamma_5 \nu_2 J.
\end{equation}
Here $\nu_1$ and $\nu_2$ are assumed to be Majorana neutrinos. Since the
weak interaction couples only to left-handed neutrinos and right-handed
antineutrinos, and Majorana particles are identical to their
anti-particles, Majorana neutrinos and Majorana antineutrinos can only be
distinguished by their spin.
The matrix elements in the observer's rest frame are given by \cite{Kim:1990km}
\begin{eqnarray}
 \label{M21}
 \left| \mathcal{M}(\nu_2 \rightarrow \nu_1) \right|^2 & = & {g_1^2
 \over 4} (A+2) + {g_2^2 \over 4} (A-2), \\
 \left| \mathcal{M}(\nu_2 \rightarrow \bar{\nu}_1) \right|^2 & = &
 {g_1^2 + g_2^2 \over 4} \left( {m_1^2+m_2^2 \over m_1 m_2} -A
 \right) = {g_1^2 + g_2^2 \over 4} \left( {1 \over x} + x - A \right),
 \label{M21anti}
\end{eqnarray}
where we have with the assumed mass hierarchy $m_2>m_1$ and
\begin{eqnarray}
 \label{MAdef}
A & \equiv & {m_1 E_2 \over m_2 E_1} + {m_2 E_1 \over m_1 E_2} = {1
\over x} {E_2 \over E_1} + x {E_1 \over E_2}, \nonumber\\
 x & \equiv & {m_2 \over m_1} \ > \ 1. \nonumber
\end{eqnarray}
The differential decay rate for the decay $\nu_2 \rightarrow
\nu_1$ is
\begin{equation}
 {d \Gamma_{2 \rightarrow 1} \over d E_1} = {m_1 m_2 \over 4 \pi} {1
 \over E_2 | \mathbf{p}_2 |} | \mathcal{M} (E_1) |^2, 
\label{dGam}
\end{equation}
with the kinematics in spherical coordinates ($z$-axis in the direction of
propagation)
\begin{equation}
 E_2 - E_1 = | \mathbf{p}_2 - \mathbf{p}_1| =
 \sqrt{\mathbf{p}_1^2 + \mathbf{p}_2^2 - 2 | \mathbf{p}_1 | |
 \mathbf{p}_2 | \cos \theta}.
 \label{KinMain}
\end{equation}
The condition $\cos \theta \leq 1$ implies for the allowed energy range of
$E_1$ that
\begin{equation}
 \label{Eborders}
\frac{E_2}{2} \left(1+{1 \over x^2} \right) - \frac{|\mathbf{p}_2|}{2}
\left( 1- {1 \over x^2} \right) \le E_1 \le \frac{E_2}{2} \left(1+{1
\over x^2} \right) + \frac{|\mathbf{p}_2|}{2} \left( 1- {1 \over x^2} \right).
\end{equation}
For relativistic neutrinos, \ie, $m_2 \ll E_2$, this reduces to
\begin{equation}
 {E_2 \over x^2} \le E_1 \le E_2.
\end{equation} 
We immediately see that for $x \simeq 1$, \ie, $\Delta
m^2 \ll m^2$, the energy shift by decay is very small. For an energy
resolution
\begin{equation}
 \Delta E_{\rm res} \gg E_2 \left( 1- {1 \over x^2} \right)
\end{equation}
the decay products are detected in the same energy bin as
the original particles.

Now, we can integrate the differential decay rate given in
\eq~(\ref{dGam}) over the range determined by \eq~(\ref{Eborders}).
Using \eqs~(\ref{M21}), (\ref{M21anti}), and (\ref{MAdef}) as well as
assuming relativistic neutrinos, \ie, $m_2 \ll E_2$, we obtain for the total
decay rates \cite{Kim:1990km}
\begin{eqnarray}
 \Gamma(\nu_2 \rightarrow \nu_1) = {m_1 m_2 \over 16 \pi E_2}
 \bigg[ & g_1^2 & \left( {x \over 2} + 2 + {2 \over x} \ln x - {2
 \over x^2} - {1 \over 2 x^3} \right) \nonumber \\
  \quad + & g_2^2 &\left( {x \over 2} - 2 + {2 \over x} \ln x + {2
 \over x^2} - {1 \over 2 x^3} \right) \bigg]
 \label{Gtot}
\end{eqnarray}
and
\begin{equation}
 \Gamma(\nu_2 \rightarrow \bar{\nu}_1) = {m_1 m_2 \over 16 \pi E_2}
       (g_1^2+g_2^2) \left( {x \over 2} - {2 \over x} \ln x - {1 
       \over 2 x^3} \right).
\end{equation}

\begin{figure}
\begin{center}
\includegraphics*[height=10cm,angle=270]{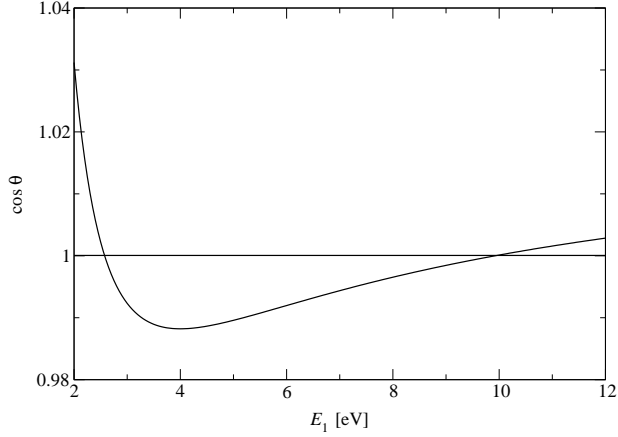}
\end{center}
\caption{\label{costhetap} The function $\cos \theta = \cos \theta
(E_1)$ plotted for the (nonrelativistic) sample data $m_1=1 \ \rm{eV}$,
$m_2=2 \ \rm{eV}$, and $E_2=10 \ \rm{eV}$.
Since $\cos \theta \leq 1$, this function is only defined below the
horizontal line.} \end{figure}

\begin{figure}
\begin{center}
\includegraphics*{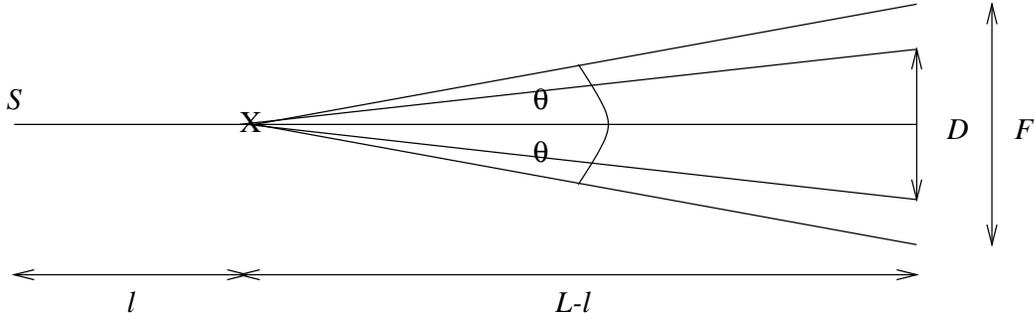}
\end{center}
\caption{\label{Beam} A neutrino beam pointing from a source $S$ to the
detection area $D$ with decay at X after traveling a distance $l$ along the
baseline $L$. The decay products of the beam cover the area $F$ at the surface
of the detector by a beam-spread of the angle $2 \theta$.} \end{figure}

\begin{table}
\caption{\label{DecayAreas} Typical energies ($E_2$),
distances ($d$), and areas ($F$) covered by decay beam-spread for
different types of neutrinos. Here it is assumed that $m =
\mathcal{O}(\rm{eV})$ and $x = \mathcal{O}(1)$.}
\begin{center}
\begin{tabular}{lccc}
Neutrino type & $E_2$/$\rm{MeV}$ & $d$/$\rm{m}$ & $F$/$\rm{m^2}$ \\ \hline
 Solar & $0.1 \sim 10$ & $\sim 10^{11}$ & $10^{10} \sim 10^{11}$ \\
 Atmospheric & $10^2 \sim 10^3$ & $10^4 \sim 10^7$ & $10^{-12} \sim
 10^{-2}$ \\
 Accelerator & $10^3 \sim 10^4$ & $10^2 \sim 10^3$ & $ 10^{-16} \sim
 10^{-12}$ \\
 Reactor & $1 \sim 10$ & $10 \sim 10^2$ & $ 10^{-12} \sim 10^{-8}$ \\
 Supernova & $\sim 10$ & $\sim 10^{21}$ & $\sim 10^{16}$ \\ \hline
 \end{tabular} 
\end{center}
\end{table}

\subsection{Visibility of decay products}
\label{ss:vis}

Solving \eq~(\ref{KinMain}) for $\cos \theta$ shows that $\cos
\theta$ as a function of $E_1$ always has a minimum at $E_{1,\min} = {2
E_2 m_1^2 \over m_1^2 + m_2^2}$ (\cf, \fig~\ref{costhetap}). We can
use this to constrain $\theta$. For relativistic neutrinos, \ie, $m_2 \ll
E_2$, we obtain
\begin{equation}
\theta_{\max} = {m_1 \over 2 E_2} \left( x^2-1 \right).
 \label{thetamax}
\end{equation}
Let us assume $m_1$ and $m_2$ to be comparable by orders of
magnitude, \ie, $x \sim 1$, which is plausible for active neutrinos
for the currently assumed mass squared differences.
For $d$, the distance to the detector, the area $F$ covered by a neutrino beam
at the detector due to a beam-spread by decay is approximately given by
\begin{equation}
 F \simeq \pi (\theta_{\max} d)^2 = \pi d^2 \left({m_1 \over 2
E_2}\right)^2 \left( x^2-1 \right)^2 = \mathcal{O} \left( \left( {d \
m_1 \over E_2} \right)^2 \right)
 \label{CovArea}
\end{equation}
for $x = \mathcal{O}(1)$.
Table~\ref{DecayAreas} shows typical values for several cases, where
neutrino beams and not point sources are assumed. For a beam-spread
at the detector (area $F$) much smaller than a usual detection area ($D$), $F
\ll D$, everything will be visible, \ie, in the cases of atmospheric,
accelerator, and reactor neutrinos.

Let us now introduce a function $\eta$ which describes the fraction
of the decay products that will arrive at the detector by geometry, \ie, does
not escape from detection by changing direction. We will
first of all consider the case of a neutrino beam, and discuss the case
of a point source later. We introduce this function in a quite general
form applicable to any decay model such as a model involving more than
two decay products. Thus, $\eta$ is a function of intrinsic decay
properties such as energy and mass of the original particle $E_{i}$, $m_{i}$,
and of the decay products $E_{j}$, $m_{j}$, as well as geometric
properties such as length of the baseline $L$, decay position $l$, and
detection area perpendicular to the beam direction $D$ (see \fig~\ref{Beam}),
\ie,
\begin{equation}
\label{gfactor}
 \eta = \eta(E_{i},m_{i},E_{j},m_{j},L,l,D) \equiv \eta_{ij}(L,l,D).
\end{equation}
Note that the intrinsic properties of the beam and the detector can be
independently folded with the transition probability, since they are not
directly affected by decay. 

Since we assume that we cannot detect particles other
than the neutrino decay product, we have to integrate over all momenta and
energies except for the angle $\theta$ (\cf, \fig~\ref{Beam}) and the energy
of the decay product $E'$. In general, these two parameters are to be
integrated over the relevant detection region and the energy bin:
\begin{equation}
 \eta_{ij}(L,l,D) = {1 \over \Gamma_{ij}^{\rm tot}} \left(
\int\limits_{E_{\min}}^{E_{\max}} \int\limits_{(\cos \theta)_D}^{1} 
\left| \frac{d^2\Gamma_{ij}}{d \cos \theta d E'} \right| d E' d \cos \theta
\right).
\label{gintegral}
\end{equation}
Here $[E_{\min},E_{\max}]$ is the energy bin range and $(\cos \theta)_D$ is
the cosine of the maximal angle, such that the decay product will still hit
the detector. Since we only have one independent parameter in Majoron decay,
the double-differential decay rate $\frac{d^2\Gamma_{ij}}{d \cos
\theta dE'}$ will reduce to ${d \Gamma \over d E'}$ times a
$\delta$-distribution determined by \eq~(\ref{KinMain}). Assuming that
the energy bin width $\Delta E_{\rm bin}$ and the angle $\theta_D$ are
very small, \ie, the differential decay rate is approximately constant
within the target region and the energy bin range, we obtain for $\eta$
\begin{equation}
 \eta_{ij}(L,l,D) \simeq {1 \over \Gamma_{ij}^{\rm tot}} \left|
\frac{d^2\Gamma_{ij}}{d \cos \theta d E'} \right|_{\cos \theta = 1, \ E'
=\bar{E}_{\rm bin}} \Delta E_{\rm bin} \ \Delta (\cos \theta)_D, 
\label{ggeneral} \end{equation}
where $\bar{E}_{\rm bin}$ is the mean energy of the target bin considered,
$\Delta E_{\rm bin}$ the energy bin width of that bin, and $\Delta (\cos
\theta)_D$ the cosine range, \ie, $\Delta (\cos \theta)_D = 1- (\cos
\theta)_D$, of the angle covered by the detector. {}From geometry
(\fig~\ref{Beam}) we can determine $\Delta (\cos \theta)_D$ for the decay
position $l$ and a spherical detection area $D$:
\begin{equation}
 2 \pi (L-l)^2 \Delta (\cos \theta)_D \simeq D.
 \label{thetageo}
\end{equation}

\begin{figure}
\begin{center}
\includegraphics*[height=10cm,angle=270]{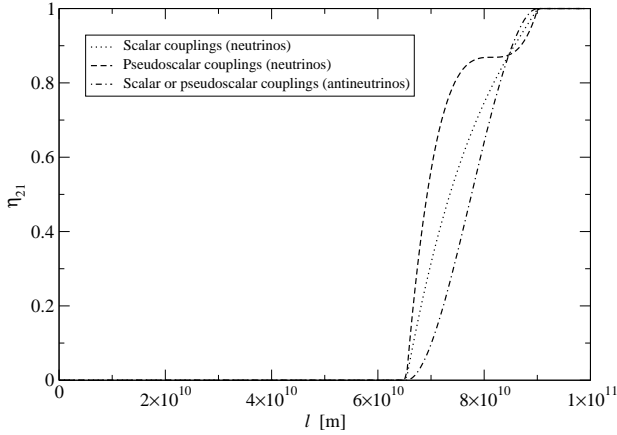}
\end{center}
\caption{\label{SolarG} The fraction $\eta_{21} = \eta_{21}(l)$ for
decay products of a solar neutrino beam hitting the detector as a 
function of the traveling distance from the surface of the Sun $l$. The
sample data used are $E_2 = 10 \ \rm{MeV}$, $m_1 = 0.02 \ \rm{eV}$, $m_2 = 0.07
\ \rm{eV}$ (\cf, \Ref \cite{Acker:1992eh}), $L = 10^{11} \ \rm{m}$, and
$D=10^5 \ \rm{m}^2$. Note that $\eta_{21}^{\rm Sun} \approx 1$ for
the Sun, because of the flux profile. }
\end{figure}

Table~\ref{DecayAreas} shows that for solar neutrinos the spread
of a beam at the detector is normally larger than the detection area.
Figure~\ref{SolarG} shows a numerical evaluation of $\eta_{21}$ for solar
neutrino sample data and a very large detection area for illustration. One
can see that in all cases the $1/(L-l)^2$-dependence dominates in the range
of about $6 \cdot 10^{10} \ \rm{m}$ to $9 \cdot 10^{10} \ \rm{m}$ over
the effects of the couplings or helicities, so that we obtain the same
qualitative behavior. However, the Sun is a radially symmetric ($4
\pi$) source. It is not a point source, because the production area is large
compared to the distance to the Earth. Equation (\ref{thetamax}) indicates
that for decay the maximal angle of deviation of the direction of flight
($\theta \approx 10^{-7}$) is much smaller than the angle under which the
Sun's core, where the neutrinos are produced, is observed from the Earth
($\theta \approx 10^{-3}$). Hence, equally
distributed small deviations of the direction of flight will already be washed
out by the flux profile. Besides that, after decay, the profile will still
remain radially symmetric. Thus, for the case of the Sun, the function
$\eta_{21}$ describing a beam can only serve as an example to show the
qualitative behavior, and we conclude that $\eta^{\rm Sun} \approx 1$.

Supernova neutrinos are different from solar neutrinos, since they come from a
far-distant point source, and hence from a well-defined direction. One
might argue that it is impossible to distinguish neutrinos produced by decay
in the direction of flight or by decay in slightly different paths due
to the radially symmetric flux profile, since decay scatters the neutrinos back
into the direction of the detector. However, these two cases may be
distinguishable for certain cases or sets of parameters because of the time
resolution (\eg, different traveling times or different oscillation lengths due
to different path lengths). Hence, supernova neutrinos might be the only
nontrivial example of partly visible Majoron decay, \ie, $\eta<1$. In general,
any far-distant beam source will show similar effects for Majoron decay. 

Eventually, we conclude that Majoron decay products are in most cases visible
to a good approximation, \ie, $\eta_{ij}(L,l,D) \approx 1$. In our
calculations, we will generalize this to $\eta_{ij}(L,l,D) \approx
\eta_{ij}(L,D)$. Hence, $\eta$ will be assumed not to depend on the distance
$l$. Moreover, for supernova neutrinos it should be a good approximation to
use a step function for $\eta_{ij}(L,l,D)$ according to \fig~\ref{SolarG}.

\subsection{Neutrino oscillations of decay products}

In this subsection, we demonstrate that neutrino decay products
in Majoron decay may oscillate. Since neutrino oscillations are
a consequence of phase coherence, several issues connected to this subject
need to be addressed for a combined treatment of neutrino decay and neutrino
oscillations. We will do that in detail in \Subsec~\ref{subsec:prop}.

In a charged current weak interaction, it is normally assumed that the phase
coherence within a superposition of states as an {\em in} state is
destroyed. This means that such a process produces a superposition of states
with random relative phase shifts, \ie, the states within
the superposition cannot interfere. Hence, cross sections referring to
the individual states have to be summed incoherently, \ie, by summation of
probabilities and not amplitudes (\cf,
\property~\ref{prop:cohincohsum} in \Subsec~\ref{subsec:prop}). This 
comes from kinematical features related to the mass differences of
accompanying leptons \cite{Smirnov:1992eg}, \ie, the coherence length is
relatively small due to the large mass differences.

In a neutral current weak interaction, the {\em out} state can be a
coherent superposition of states. One example is $Z^{0}$ decay,
producing a coherent superposition of different neutrino and antineutrino mass
eigenstates \cite{Smirnov:1992eg}
\begin{equation}
 | \nu_{Z} \rangle \equiv {1 \over \sqrt{3}} \left( | \bar{\nu}_1 \rangle |
\nu_1 \rangle + | \bar{\nu}_2 \rangle | \nu_2 \rangle + | \bar{\nu}_3
\rangle | \nu_3 \rangle \right) 
\end{equation}
by an interaction Lagrangian
\begin{equation}
 \label{ZLagrangian}
 \mathcal{L}_{\rm int} = i {g \over 2} \left( \sum\limits_{i=1}^{3}
 \bar{\nu}_{i L} \gamma^{\mu} \nu_{i L} \right) Z^0_{\mu}.
\end{equation}

Majoron decay resembles a neutral current weak interaction in the sense of a
Lagrangian similar to \eq~(\ref{ZLagrangian}). It couples a
superposition of different mass eigenstates to a single particle,
so that all eigenstates and their wave packets, respectively, experience the
same energy shift $\Delta E$ transferred to the massless boson. {}From the wave
packet treatment of neutrino oscillations we know that within a certain
coherence length, the mass differences of neutrinos do not destroy the
coherence of the mass eigenstates, \ie, neutrino oscillations. Hence, since in
Majoron decay the energies of the mass eigenstates are all shifted by the same
amount $\Delta E$ in decay, the mass differences will not be able to destroy
the coherence of {\em out} states in this case either\footnote{Later in this
paper, a more detailed proof of this analogy applied to our framework will be
given in \property~\ref{prop:noscdp} in \Subsec~\ref{subsec:prop}.}. Thus, the
decay products can oscillate.

Using a Majoron-like decay interaction Lagrangian, we know that to first
order of the $S$-matrix expansion we can write
\begin{equation} d \Gamma(in
\rightarrow out) \propto \left| \left< out \right| \int\limits_{-
\infty}^{\infty} d^4 x \underset{i \neq j}{\sum_i \sum_j} g_{ij}
\bar{\nu}_{j} \nu_{i} J \left| in \right> \right|^2.
\label{Lint}
\end{equation}
This expression implies that the interaction Lagrangian destroys the
incoming superposition completely,
and creates a superposition of outgoing mass eigenstates with coefficients
determined by the coupling constants and the coefficients of the incoming
superposition. The outgoing states have a fixed relative initial phase such as
the outgoing mass eigenstates in $Z^0$ decay \cite{Smirnov:1992eg}. In order
to observe neutrino oscillations of decay products, we need to have at least
two different mass eigenstates in the {\em out} state superposition. We
immediately see from \eq~(\ref{Lint}) that this implies that at least two of
the $g_{ij}$'s, say $g_{i_1j_1}$ and $g_{i_2j_2}$, have to be non-zero for the
corresponding states $\left| \nu_{i_1} \right>$ and $\left| \nu_{i_2} \right>$
with non-vanishing {\em in} state amplitudes $\nu_{i_1}$ and $\nu_{i_2}$.

\section{Operators and properties}
\label{sec:oap}

In this section, we will introduce effective decay operators for a
model-indepen\-dent combined treatment of neutrino decay and neutrino
oscillations. Decay can be described by decay rates $\Gamma_{ij}
\propto g_{ij}^2$, which are calculated from the
corresponding Feynman diagrams. We will use such an approach to effectively
integrate neutrino decay into neutrino oscillation scenarios. For this we do
not need details of the specific model as long as
neutrino decay can be described by a generic effective Lagrangian destroying
one neutrino mass eigenstate and creating another one, \ie,
$\mathcal{L}_{\rm int} \propto g_{ij} \bar{\nu}_j \nu_i$. This also
includes multi-particle decay into decay products other than
neutrinos, where all momenta and energies of 
the undetected decay products are to be integrated over in the decay
rates\footnote{For decay models involving more than one outgoing neutrino we
could use a similar framework with slightly different properties. For
Majoron-like models higher-order processes are assumed to be suppressed,
because the coupling constants are considered to be small.}. We will show that
our operator framework satisfies the usual properties of the $S$-matrix
approach and implements the correct phase relations. In certain cases, we will
demonstrate this for Majoron decay.

\subsection{Operators}

Let us introduce decay and propagation operators in terms of
creation and annihilation operators, $\hat{a}^{\dagger}$ and $\hat{a}$,
respectively, as well as calculation rules for the transition probabilities.
The symbols introduced in the operators will be explained thereafter. In the
next subsection, we will show how the operators satisfy the expected
properties. We define three operators:

\begin{definition}[Disappearance operator]
$\mathcal{D}_{-}$ is the transition operator generally known as the
``decay operator''.
Effectively, it returns the amplitude for an undecayed state remaining
undecayed after traveling a distance $l$ along the baseline $L$:
\begin{equation}
 \label{DMinus}
 \mathcal{D}_{-}(l) = \sum_{i} \exp \left( - \frac{\alpha_{i} l }{2 E_{i}}
\right) \hat{a}_{i}^{\dagger} \hat{a}_{i}.
\end{equation}
\end{definition}

\begin{definition}[Appearance operator]
$\mathcal{D}_{+}$ is the differential transition operator, which destroys an
{\bf in} state and creates an {\bf out} state in $[l,l+dl]$ along the
baseline $L$, \ie, a new state ``appears'':
\begin{equation}
 \label{DPlus}
 \mathcal{D}_{+}(l,L) = \underset{i \neq j}{\sum\limits_i \sum\limits_j}
\sqrt{\frac{\alpha_{ij}}{E_{i}}} \ \sqrt{\eta_{ij}(L,l,D)} \,
 \hat{a}_{j}^{\dagger} \hat{a}_{i} e^{i \xi}.
\end{equation}
The phase $\xi$ is a random phase taking into account the phase shift by
additional (not measured) particles produced in the decay, such as Majorons
(\cf, \property~\ref{prop:phcohinout} in \Subsec~\ref{subsec:prop}).
The probability density, which is the square of the amplitude, will
have to be integrated over $l$.
\end{definition}

\begin{definition}[Propagation operator]
$\mathcal{E}$ is the operator propagating a state a distance $l$ along the
baseline $L$\footnote{Here we are using the
non-relativistic operator, which is different from the
relativistic one only by an overall phase factor. This phase factor will
cancel in the calculation of transition probabilities. For justification of
this approximation see \Ref \cite{Giunti:2000kw}.}:
\begin{equation} 
\label{Evol} \mathcal{E}(l) = \sum_{i} \exp \left( - i E_{i} l \right) 
\hat{a}_{i}^{\dagger} \hat{a}_{i}.
\end{equation}
\end{definition}

In these definitions, $\eta_{ij}(L,l,D)$ is the geometrical function introduced
in \eq~(\ref{gfactor}) and defined in \eq~(\ref{gintegral}). It describes the
fraction of decay products which will still arrive at the detector, \ie, is not
redirected from the detector by decay. The decay rate is defined as
$\alpha_{ij} \equiv \frac{m_{i}}{\tau^{0}_{ij}}$ for $i \rightarrow j$
decay, and $\tau^{0}_{ij}$ refers to the rest frame lifetime for that
decay channel. Time dilation by the factor $\gamma_i=E_i/m_i$ implies that
\begin{equation}
\Gamma_{ij}^{\rm observer} = \Gamma_{ij}^{0} \gamma_i^{-1} =
{1 \over \tau_{ij}^0} \frac{m_i}{E_i} =
{\alpha_{ij} \over E_i}.
\end{equation}
The rate $\alpha_{i} \equiv \sum_{j} \alpha_{ij}$ is the overall decay
rate of the state $i$, so that $B \equiv \frac{\alpha_{ij}}{\alpha_{i}}$ is the
branching ratio for $i \rightarrow j$ decay. The factor of two in the
exponent of $\mathcal{D}_{-}$ and the square root
in $\mathcal{D}_{+}$ comes from the fact that, besides some phase factors,
amplitudes effectively behave like square roots of particles. However,
as we will see in the next subsection, the
above operators together with the definitions of the transition probabilities
also implement the correct phase relations.
Let us now define how to calculate the transition
probabilities. This is to be done by successive application of decay
and propagation operators:

\begin{definition}[Calculation of transition probabilities]
\label{def:transprob}
The transition probability $P(\nu_{\alpha} \rightarrow \nu_{\beta})^n=P_{\alpha
\beta}^n$ by {\bf exactly} $n$ intermediate decays is for $n=0$ given by
\begin{equation}
 \label{pdecay0}
 P_{\alpha \beta}^0 = \left| \left< \nu_{\beta} |
\mathcal{E}(L) \mathcal{D}_{-}(L)
 | \nu_{\alpha} \right> \right|^2,
\end{equation}
and for $n>0$ given by
\begin{eqnarray}
P_{\alpha \beta}^n &=& \int\limits_{l_1=0}^{L} \cdots
\int\limits_{l_n=l_{n-1}}^{L} | \langle \nu_{\beta} |
\mathcal{E}(L-l) \mathcal{D}_{-}(L-l) \nonumber\\
&\times& \prod\limits_{i=1}^{n} \Big\{ \mathcal{D}_{+}(l_i,L)
\mathcal{E}(l_i) \mathcal{D}_{-}(l_i) \Big\} | \nu_{\alpha} \rangle |^2
\prod\limits_{i=1}^{n} dl_i.
\label{pdecay}
\end{eqnarray}
Here $l \equiv \sum_{i=1}^{n} l_i$. The order of $\mathcal{D}_{-}$ and
$\mathcal{E}$ is arbitrary, since these
operators commute as we will see below, in \eq~(\ref{comm1}).

In many cases, the total transition probability can be calculated as
\begin{equation}
\label{rdecay}
P_{\alpha \beta} =
\sum\limits_{n=0}^{N_{\max}} P_{\alpha \beta}^n, 
\end{equation}
where $N_{\max}$ denotes the maximal number of decays possible, depending on
the decay model. The individual
terms in \eq~(\ref{rdecay}) add up only if the particles produced by a
different number of decays cannot be distinguished, such as by energy
resolution, energy threshold or signature (flavor or spin). Depending on the
problem, the terms may have to be split appropriately in general\footnote{For
instance, we may even have two different $P^2_{\alpha \beta}$ coming from
$\nu_i \rightarrow \bar{\nu}_j \rightarrow \nu_k$ and $\nu_i \rightarrow \nu_j
\rightarrow \nu_k$ described by the appropriate decay rates. Another example
is energy binning, where one $P^i_{\alpha \beta}$ may have to be split into
fractions of different energy ranges to be calculated from the differential
decay rate, such as \eq~(\ref{dGam}) for Majoron decay.}.
\end{definition}

Finally, we can introduce two simplified definitions for {\em invisible} as
well as stable or almost stable {\em visible} decay products:

\begin{definition}[{\em Invisible} decay]
For {\bf invisible} decay products we can, in principle, only observe the
undecayed particles themselves. Hence, we do not admit appearance operators,
\ie,
\begin{equation}
 P_{\alpha \beta}^{\rm invisible} = P_{\alpha \beta}^0.
 \label{pinvisible}
\end{equation}
\end{definition}

\begin{definition}[Approximations for {\em visible} decay]
Considering maximal\\ one decay or lifetimes longer than $Lm \over E$,
we can neglect repeated decay terms ($n>1$) and approximate the total
transition probability in \eq~(\ref{rdecay}) by
\begin{equation}
 P_{\alpha \beta} \simeq P_{\alpha \beta}^0 + P_{\alpha \beta}^1 = P_{\alpha
\beta}^{\rm invisible} + P_{\alpha \beta}^{\rm appearance},
\label{totapprox1}
\end{equation}
where $P_{\alpha \beta}^{\rm appearance}$ describes the ``appearance'' of
new particles by decay.
{}From \eq~(\ref{pdecay}) we can read off the first correction term $P_{\alpha
\beta}^1$ for visible decay in addition to $P_{\alpha \beta}^{\rm invisible}$:
\begin{equation}
 \label{papprox}
 P_{\alpha \beta}^1 = \int\limits_{l=0}^{L} \left| \left<
\nu_{\beta} | \mathcal{E}(L-l) \mathcal{D}_{-}(L-l) \mathcal{D}_{+}(l,L)
\mathcal{E}(l) \mathcal{D}_{-}(l) | \nu_{\alpha}
\right> \right|^2 dl.
\end{equation}
\end{definition}

\subsection{Properties}
\label{subsec:prop}

In this subsection, we want to make the most important properties of our
operator framework more transparent. Since we deal with decay of
states instead of decay of particles, we need to show that this
describes particle decay correctly while preserving interference
properties for states.

\begin{Prop}[Creation and annihilation operators]
The effective creation and annihilation operators, $\hat{a}^{\dagger}$
and $\hat{a}$, respectively, are consistent with the $S$-matrix approach.
\end{Prop}
The operator $\hat{N}_i \equiv \hat{a}_i^{\dagger} \hat{a}_i$ in
\eqs~(\ref{DMinus}) and (\ref{Evol}) is as usual the occupation number
operator, whereas the operator $\hat{T}_{ij} \equiv
\hat{a}_j^{\dagger} \hat{a}_i$ in \eq~(\ref{DPlus}) is an effective
transition operator coming from inserting the fermionic neutrino field
expansions into an equation like \eq~(\ref{Lint}). Here $i$ and $j$
refer to all quantum numbers, such as spin and momentum. Since we
assume that conceptual issues, such as detection in a different energy
bin or decay into an antiparticle, are dealt with by splitting
\eq~(\ref{rdecay}) appropriately (\cf, \Def~\ref{def:transprob}), we 
do not need to take care of different quantum numbers here.

\begin{Prop}[Commutation relations]
The operators $\mathcal{D}_{-}$ and $\mathcal{E}$ satisfy the commutation
relation
\begin{equation}
 \label{comm1}
 \left[ \mathcal{D}_{-},\mathcal{E} \right] = 0,
\end{equation}
whereas in general
\begin{equation}
 \label{comm2}
 \left[ \mathcal{D}_{-},\mathcal{D}_{+} \right] \neq 0, \quad 
 \left[ \mathcal{D}_{+},\mathcal{E} \right] \neq 0.
\end{equation} 
\end{Prop}
These relations can be shown by direct calculation of the commutators, using
the fermionic anticommutation relations $\{ \hat{a}_i, \hat{a}_j^{\dagger}
\} = \delta_{ij}$.

One can also derive \eq~(\ref{comm1}) by introducing the notion of a complex
mass square, \ie, 
$$
 \widetilde{m}_i^2 \equiv m_i^2 - i \alpha_i.
$$
For $\tilde{\mathcal{E}}(l) = \sum_{i} \exp \left( - i \tilde{E}_{i} l
\right) \hat{a}_{i}^{\dagger} \hat{a}_{i}$, where $\tilde{E}_i \simeq p +
{\widetilde{m}_i^2 \over 2 E_i}$, we can read off from \eqs~(\ref{DMinus}) and
(\ref{Evol}) that
$$
\mathcal{D}_{-}(l) \mathcal{E}(l) =
\tilde{\mathcal{E}}(l) = \mathcal{E}(l) \mathcal{D}_{-}(l).
$$
This means that the disappearance and propagation operators can be combined by
the introduction of complex mass squares. Therefore, the commutator reduces to
the commutator of two scalars, \ie, the exponentials of the real and imaginary
parts of the mass square.

\begin{Prop}[Decay into different channels]
Our description for the decay of states corresponds to a description for
the decay of particles, which is given by the differential equation system
\begin{equation}
 \label{DSdecay}
 {d \over dt} N_i = - \sum_{j, \ j \neq i} \Gamma_{ij} N_i + \sum_{j, \ j \neq
i} \Gamma_{ji} N_{j}.
\end{equation}
Here $\Gamma_{ij}$ is the rest frame decay rate for the particle decay $i
\rightarrow j$ and $N_i$ is the number of particles $i$.
{}From this we obtain for the disappearance of a particle $N_{i,0}=1$,
$N_{j,0}=0$ for all $j \neq i$, and $\Gamma_{i} \equiv \sum_{j, j \neq i}
\Gamma_{ij}$ \begin{equation} 
 \label{DecaySolutionD} 
 N_i(t) = e^{- \Gamma_{i} t},
\end{equation}
and for the appearance of a stable decay product $N_{i,0}=0$, $N_{j,0}=1$ for
all $j \neq i$, and $\Gamma_i=0$
\begin{equation}
\label{DecaySolutionA}
N_i(t) = \sum_{j, \ j \neq i} \frac{\Gamma_{ji}}{\Gamma_{j}} \left( 1- e^{-
\Gamma_{j} t} \right).
\end{equation}
\end{Prop}

Neglecting propagation operators\footnote{In this case, we are
not interested in neutrino oscillation probabilities and phase factors
will cancel anyway when squaring the probability amplitudes.}, we obtain from
\eq~(\ref{pdecay0}) for the disappearance probability of a mass eigenstate
\begin{equation}
P_{ii} = \left| \left< \nu_i | \mathcal{D}_{-}(l)| \nu_i \right>
\right|^2 = e^{- \frac{\alpha_i}{E_i} L},
\label{praw0}
\end{equation}
which is identical to the disappearance probability
in \eq~(\ref{DecaySolutionD}) if $\Gamma_i = {\alpha_i \over E_i}$, which we
have in the observer's rest frame.

Neglecting propagation operators, we obtain from \eq~(\ref{papprox})
for the appearance of a stable decay
product ($\alpha_j=0$) for exactly one decay
\begin{equation}
P_{ij} = \int_{l=0}^L \left| \left< \nu_j | \mathcal{D}_{+}(l,L) 
\mathcal{D}_{-}(l)| \nu_i \right> \right|^2 dl = 
\eta \frac{\alpha_{ij}}{\alpha_i} \left( 1 - e^{- \frac{\alpha_i}{E_i}
L} \right),
\label{praw}
\end{equation}
which is identical to the appearance probability
in \eq~(\ref{DecaySolutionA}) if $\eta=1$ (everything detected) and $\Gamma_i =
{\alpha_i \over E_i}$, which we again have in the observer's rest frame.

In general, \eqs~(\ref{pdecay}) and (\ref{DSdecay}) obey the
same structure, since transitions among states are described by differential
rates as in a Markov chain model. 

\begin{Prop}[Coherent and incoherent summation of amplitudes]
\label{prop:cohincohsum}
In quantum field theory, {\bf coherent} and {\bf incoherent} summation are
often distinguished in the calculation of Feynman diagrams
\cite{Grimus:1998uh}. The notion {\bf coherent} refers to the
summation of amplitudes
$$
 d \Gamma \propto \left| \sum\limits_{i} A_i \right|^2
$$
and {\bf incoherent} to the summation of squares of
amplitudes, \ie, summation of probabilities
$$
 d \Gamma \propto \sum\limits_{i} \left|A_i \right|^2.
$$
Basically, {\bf coherent} summation applies to processes
which are, in principle, indistinguishable and {\bf incoherent}
summation to
processes which are, in principle, distinguishable.
Our formalism implements the proper application of
{\bf coherent} or {\bf incoherent} summation.
\end{Prop}
To show this, we split the statement into more specific parts in what follows.

\begin{Prop}[Incoherent summation and repeated decay]
\label{prop:rdecay}
Repeated decay can be described by the incoherent summation in
\eq~(\ref{rdecay}).
\end{Prop}
The differential decay rate in terms of the $S$-matrix is
\begin{equation}
 d \Gamma(in \rightarrow out) \propto \left| \left< out \left| S \right| in
\right> \right|^2,
 \label{pinout}
\end{equation}
where $S$ can be expanded as usually as \cite{MandlShaw}
\begin{equation}
 \label{smatrix}
 S = \sum\limits_{n=0}^{\infty} {i^n \over n!}
\int\limits_{-\infty}^{\infty} \dots
\int\limits_{-\infty}^{\infty} d^4 x_1 \dots d^4 x_n T \left\{
\mathcal{L}_{\rm int}(x_1) \dots \mathcal{L}_{\rm int}(x_n) \right\}.
\end{equation}
Thus, the {\em in} and {\em out} states select the corresponding $S$-matrix
expansion terms, which will evaluate to non-zero results. For a
different number of
decays we obtain different decay products characterized by different quantum
numbers. Therefore, different $S$-matrix expansion terms will be selected in
\eq~(\ref{pinout}). This means that Feynman diagrams of different orders
cannot interfere, even if we do not distinguish them by the measured decay
products (often only the neutrinos but no other particles, such as
Majorons, are measured). Hence, decay implies the principal
possibility to measure the decay products and, thus, in terms of
quantum mechanics acts as a measurement. Since repeated decay corresponds to
higher order Feynman diagrams in quantum field theory, incoherent summation
properly describes it.

\begin{Prop}[On-shell propagation in between decays]
\label{prop:onshell}
It is shown in \Ref \cite{Grimus:1999zp} that
Feynman propagators for virtual particles reduce to on-shell propagators for
macroscopic distances\footnote{In this reference, this was explicitely shown
for the creation and detection processes.}. Thus, individual decays in repeated
decay can be treated independently without interference. More precisely,
Feynman diagrams of the same order do not produce interference terms, because
they can (in principle) be distinguished due to the large separation
of vertices.
\end{Prop}

Equation (\ref{rdecay}) together with \eq~(\ref{pdecay}) imply that
decays are treated separately with decay rates calculated as for independent
processes.

\begin{Prop}[Phase coherence between {\em in} and {\em out}
neutrinos]
\label{prop:phcohinout}
Since at least a third unmeasured
particle, \eg, the Majoron, is involved, the overall phase coherence
between {\bf in} and {\bf out} neutrinos is destroyed.
\end{Prop}
The random phase $\xi$ in \eq~(\ref{DPlus}) takes care of this. Since
it does not appear in $P_{\alpha \beta}^0$ and cancels in
$P_{\alpha \beta}^1$, it is only relevant for repeated decays for
$n>1$. It also ensures that terms from different individual decays do
not produce interference terms, \ie, Feynman diagrams of the same
order are summed incoherently in this case (\cf,
\property~\ref{prop:onshell}).

\begin{Prop}[Decay model and causality]
The decay model used in \eqs~(\ref{pdecay0}) and (\ref{pdecay})
properly implements causality. 
\end{Prop}
The operator $\mathcal{D}_{-}$ gives the probability amplitude
for a state remaining undecayed until the position $l$. The operator
$\mathcal{D}_{+}$ gives the conditional probability amplitude for
decay in the interval 
$[l,l+dl]$ for an undecayed state. Successive
alternating applications of $\mathcal{D}_{-}$ and $\mathcal{D}_{+}$ give the
(conditional) probability amplitude for the decision path.
This probability needs to be integrated to obtain the average over
all possible decay positions $l$ in a causal order, which is taken into
account by the integration limits, \ie, a particle cannot decay before it
even exists. We are integrating probability amplitudes for different decay
positions incoherently, taking into account that undecayed particles cannot
interfere with decay products.

\begin{Prop}[Neutrino oscillations of decay products]
\label{prop:noscdp}
The appearance operator $\mathcal{D}_{+}$ gives the proper relative phase
in a superposition of mass eigenstates as an {\bf out} state.
Kinematics induced by different masses does not result in destruction
of coherence among the superimposed states such as in charged current weak
interactions.
\end{Prop}

The decay rate is to first order of the $S$-matrix expansion as
usually defined as
\cite{MandlShaw}
\begin{equation}
 \Gamma ( in \rightarrow out ) = {1 \over T} \left| \langle out | \int
d^4 x \ \mathcal{L}_{\rm int} | in \rangle \right|^2.
\end{equation}
Inserting a Majoron-like interaction Lagrangian (\eg, \eq~(\ref{int2})) with
the corresponding field expansions yields for the decay rate from one flavor
$\left| \nu_{\alpha} \right> = \sum_i U_{\alpha i}^* \left| \nu_i \right>$ into
another one $\left| \nu_{\beta} \right> = \sum_j U_{\beta j}^* \left| \nu_j
\right>$
\begin{equation}
 \label{GFlavor} 
 \Gamma_{\alpha \beta} = {1 \over T} \left| \langle \nu_{\beta} | \int
d^4 x \ \mathcal{L}_{\rm int} | \nu_{\alpha} \rangle \right|^2
 = {1 \over T} \left| \sum\limits_{i} \sum\limits_{j} U_{\alpha i}^* U_{\beta
j} F_{ij} \right|^2, \end{equation}
and from one mass eigenstate $\left| \nu_i \right>$ into another one
 $\left| \nu_j \right>$
\begin{equation}
 \label{GMass}
 \Gamma_{ij} = {1 \over T} \left| \langle \nu_j | \int d^4 x
\ \mathcal{L}_{\rm int} | \nu_i \rangle \right|^2 = {1 \over T} \left| 
F_{ij} \right|^2. \end{equation}
Here the $F_{ij}$'s refer to transition functions depending on the
properties of the states $i$ and $j$ as well as the Lagrangian and the used
field expansions. Without loss of generality, for Dirac field
expansions of fermions and a Klein--Gordon field expansion of the boson, we
obtain for Lagrangians of the Majoron decay type
\begin{eqnarray}
 F_{ij} & = & \langle \nu_j | \int d^4 x \ \mathcal{L}_{\rm int} |
\nu_i \rangle \nonumber \\
& = &N_{ij} \int d^4 x e^{i p_J x} \mathcal{A} \left(
\overline{u^j} ( \mathbf{p}_2 ) e^{i p_2 x } \right) g_{ij} \mathcal{B}
\left( u^i ( \mathbf{p}_1 ) e^{-i p_1 x } \right), 
 \label{deff}
\end{eqnarray}
where $N_{ij}$ are normalization factors depending on energy, mass, and
volume and $\mathcal{A}$ and $\mathcal{B}$ are some operators depending on the
Lagrangian, \eg, combinations of charge conjugation and chiral operators.
By introducing $\phi_{ij}$ as the phase factor of $F_{ij}= \left|
F_{ij} \right| e^{i \phi_{ij}}$ determined by \eq~(\ref{deff}), we can
read off the following relation from \eq~(\ref{GMass}):
\begin{equation}
 \label{RGamma}
 \sqrt{\Gamma_{ij}} e^{i \phi_{ij}} = {1 \over \sqrt{T}} \left| F_{ij} \right|
e^{i \phi_{ij}} = {1 \over \sqrt{T}} F_{ij}.
\end{equation}
As we will show below, for
relativistic neutrinos $\phi_{ij} \approx \phi$, \ie, $\phi_{ij}$ is
independent of the indices $i$ and $j$.

Without loss of generality, in our framework the rest frame transition
probability between two flavor eigenstates for $\eta_{ij}=1$ (\ie,
everything is detected) is given by
\begin{equation}
 P_{\alpha \beta} = \left| \langle \nu_{\beta} | \mathcal{D}_{+} |
\nu_{\alpha} \rangle \right|^2 T = \left| \langle \nu_{\beta} |
\sum_{i} \sum_{j} \sqrt{\Gamma_{ij}}
\hat{a}_{j}^{\dagger} \hat{a}_i | \nu_{\alpha} \rangle \right|^2 T.
\label{eq:PabT}
\end{equation}
This can be obtained from \eq~(\ref{pdecay}) for an infinitesimally
short baseline ($dl = T$) by using $\mathcal{D}_{-} \approx
\mathcal{E} \approx 1$ to zeroth order in the expansion of the
exponentials as well as $\Gamma_{ij} = {\alpha_{ij} \over
E_i}$. Applying \eq~(\ref{RGamma}) and the expansions of the flavor
eigenstates, we finally obtain from \eq~(\ref{eq:PabT})
\begin{equation}
 \Gamma_{\alpha \beta} = {P_{\alpha \beta} \over T} = \left|
\sum\limits_{i} \sum\limits_{j} U_{\alpha i}^* U_{\beta j} {1 \over
\sqrt{T}} | F_{ij} | \right|^2 = {1 \over T} \left| \sum\limits_{i}
\sum\limits_j U_{\alpha i}^* U_{\beta j} | F_{ij} | \right|^2.
\end{equation}
This is equivalent to \eq~(\ref{GFlavor}) if $\phi_{ij} = \phi$ is
independent of the indices $i$ and $j$. In this case, the appearance
operator is identical to the expression derived from the $S$-matrix
expansion and the different masses do not destroy phase coherence in a
superposition of states.

It remains to be shown that $\phi_{ij}$ is independent of the indices $i$
and $j$ for relativistic neutrinos. Since $\phi_{ij} = \arg F_{ij}$, any real
factor in $F_{ij}$ such as, for instance,
a normalization factor, does not affect $\phi_{ij}$. We conclude from
\eq~(\ref{deff}) that for real coupling
constants the $i$ and $j$ dependence of $\phi_{ij}$ can only be influenced by
the exponentials and the spinors. By spatial
integration, the exponentials are transformed into
$\delta$-distributions implying momentum conservation and, hence,
cannot affect the phase. Without loss of generality, the
spinors can be written in the Dirac representation as \cite{MandlShaw}
\begin{equation}
u_{\uparrow}(\mathbf{p}) = N_1 \left( \begin{array}{c} 1 \\ 0 \\ N_2
p^3 \\ N_2 (p^1 + i p^2) \end{array} \right), \quad
u_{\downarrow}(\mathbf{p}) = N_1 \left( \begin{array}{c} 0 \\ 1 \\ N_2 (p^1
- i p^2) \\ - N_2 p^3 \end{array} \right),
\end{equation}
where $N_1$ and $N_2$ are (real) normalization factors. For relativistic
neutrinos we can assume the three momentum ${\bf p}$ to be identical for all
mass eigenstates and absorb the mass dependence of the
four-momentum in a correction of the energy $E_i \simeq p + {m_i^2 \over 2
p}$. Then, they do not affect $\phi_{ij}$ in this approximation,
because the phases of these spinors only depend on the
common three-momentum. Hence, $\phi_{ij}$ is independent of the indices
$i$ and $j$.

\section{Invisible decay}
\label{sec:invis}

In this section, we will treat {\em invisible} decay with our operator
framework, \ie, decay into particles, such as sterile decoupled
neutrinos, which are in
principle unobservable. In fact, this is the most often assumed neutrino decay
scenario.

\begin{figure}[htb]
\begin{center}
\includegraphics*[height=10cm]{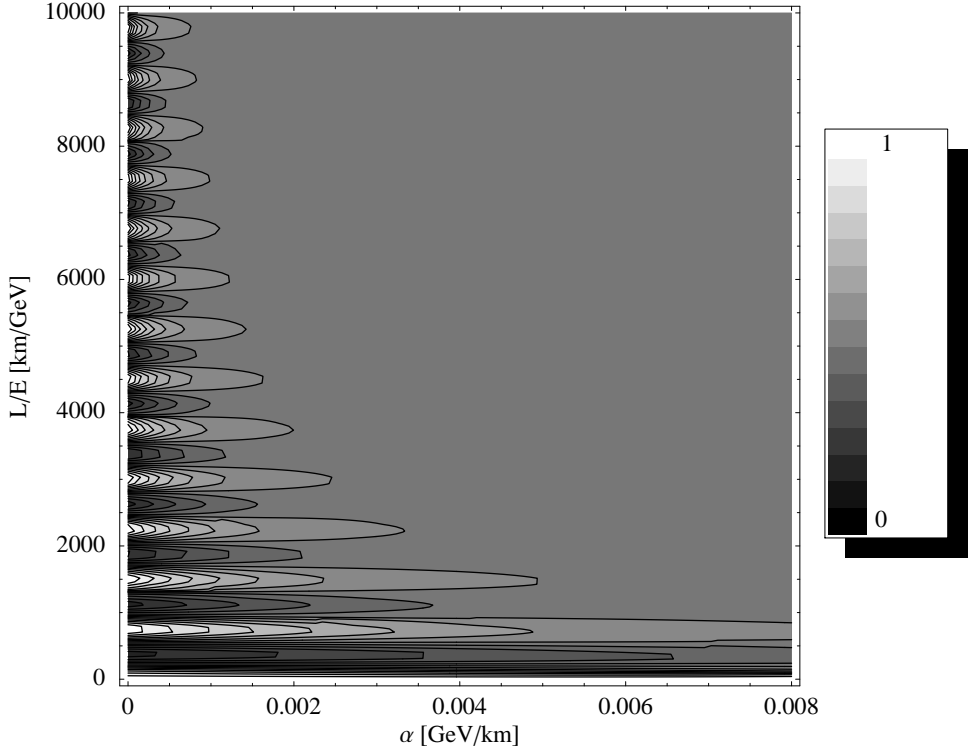}
\end{center}
\caption{\label{DecayOsc} The survival probability $P_{\mu \mu}$ as a function
of $\alpha$ and $L/E$ for $\cos^2 \theta_{23} = 0.30$ and $\Delta
m_{32}^2 = 3.3 \cdot 10^{-3} \, {\rm eV}^2$.} \end{figure}

\subsection{Transition probability for invisible decay}

The transition probability for invisible decay is calculated
according to \eqs~(\ref{pdecay0}) and (\ref{pinvisible}). Since the operator
$\mathcal{D}_{-}$ only gives real factors, the calculation for relativistic
neutrinos is quite straightforward and similar to the one for the general
neutrino oscillation formula.
We use $\left| \nu_{\alpha} \right> = \sum_{i} U_{\alpha i}^{*} \left| \nu_i
\right>$ and define the following quantities:
\begin{eqnarray}
\label{transprobJ}
J_{ij}^{\alpha \beta} & \equiv & U_{\alpha i} U^{*}_{\alpha j} U^{*}_{\beta i}
U_{\beta j}, \\
 \Delta_{ij} & \equiv & \frac{\Delta m^2_{ij} L}{4 E} \equiv \frac{ \left(
m^2_{i}- m^2_{j} \right) L}{4 E},
 \label{defdelta}
\\
 \Gamma_{ij} & \equiv & \frac{m_{i} L
}{2 \tau_{0,i} E_{i}} + \frac{m_{j} L}{2 \tau_{0,j} E_{j}} \simeq
 \left( \frac{m_{i}
}{\tau_{0,i}} + \frac{m_{j}}{\tau_{0,j}} \right) \frac{L}{2 E} \equiv 
 \left( \alpha_{i}+\alpha_{j} \right) \frac{L}{2 E}.
 \label{defgamma}
\end{eqnarray}
Here $E_i \simeq E$ has been used in the decay terms,
because for decay the zeroth order terms in $E_i \simeq p +
\frac{m_i^2}{2 p}$ do not cancel, and therefore, the higher order terms can be
neglected\footnote{For neutrino oscillations the zeroth order terms cancel due
to multiplication with their complex conjugates in the calculation of the
transition probability.}. For more details on this kind
of approximations see \Ref \cite{Giunti:2000kw}.

After some calculations, not to be shown here, we finally obtain
\begin{eqnarray}
P_{\alpha \beta}^{\rm invisible} & = & \underbrace {
\underbrace{\sum\limits_i \sum\limits_j \Re J_{ij}^{\alpha \beta} e^{-
\Gamma_{ij}}}_{P_{\rm pure \ decay}} - 4 \; \underset{i>j}{\sum\limits_i 
\sum\limits_j} 
\Re J_{ij}^{\alpha \beta} \sin^2 \Delta_{ij} e^{-\Gamma_{ij}}}_{P_{\rm
CP \, conserving}} \nonumber \\ & - & \underbrace{ 2 \;
\underset{i>j}{\sum\limits_i \sum\limits_j} \Im J_{ij}^{\alpha \beta} \sin 2
\Delta_{ij} e^{- \Gamma_{ij}}}_{P_{\rm CP \, violating}}.
\label{transprobdecay} \end{eqnarray}
In our definition, {\em pure decay} admits neutrino flavor mixing, but no
neutrino oscillations ($\Delta_{ij} = 0$ for all indices $i$ and $j$).

\subsection{Example: Application to atmospheric neutrinos}
\label{ex:appat1}

In \Ref \cite{Barger:1999bg}, neutrino decay was considered as an
alternative to atmospheric neutrino oscillations\footnote{In this paper, we
use the standard parameterization for the active neutrino mixing
matrix \cite{Groom:2000in}
$$
U_{3 \times 3} = \left( \begin{array}{ccc}
c_{12} c_{13} & s_{12} c_{13} & s_{13} e^{- i \delta} \\
-s_{12} c_{23} - c_{12} s_{23} s_{13} e^{i \delta} & c_{12} c_{23} -
 s_{12} s_{23} s_{13} e^{i \delta} & s_{23} c_{13} \\ 
 s_{12} s_{23} - c_{12} c_{23} s_{13}
e^{i \delta} & -c_{12} s_{23} - s_{12} c_{23} s_{13} e^{i \delta} & c_{23}
c_{13} 
 \end{array}
 \right),
$$
 where $s_{ij} \equiv \sin \theta_{ij}$ and $c_{ij} \equiv \cos \theta_{ij}$
and where $\theta_{ij}$ are the vacuum mixing angles. }. Assuming $\Delta
m_{21}^2=0$, $\theta_{12}=\theta_{13}=0$, and the CP-phase $\delta=0$, we
obtain from \eq~(\ref{transprobdecay}) for decay of $\nu_{2}$ into a sterile
decoupled neutrino with the rate $\alpha \equiv \alpha_2$ the same transition
probabilities as in \Ref \cite{Barger:1999bg}:
For {\em pure decay} ($\Delta m_{32}^2=0$) we have
\begin{equation}
 P_{\mu \mu} = \left( c_{23}^2 e^{-\frac{\alpha L}{2E}} + s_{23}^2 \right)^2 ,
\end{equation}
and for {\em decay and oscillations}
\begin{equation}
 \label{atmosc}
 P_{\mu \mu} = s_{23}^4 + c_{23}^4 e^{-\frac{\alpha L}{E}} + 2 c_{23}^2
s_{23}^2 \cos \left( \frac{\Delta m_{32}^2 L}{2E} \right) e^{-\frac{\alpha
L}{2E}}.
\end{equation}
The authors of \Ref \cite{Barger:1999bg} fitted Super-Kamiokande data
in the case of decay and oscillations with negligible $\Delta
m_{32}^2$ finding $\alpha = {1 \over 63} \rm{GeV
\over km}$ and $\cos^2 \theta \equiv \cos^2 \theta_{23} = 0.30$. Pure
decay without neutrino flavor mixing, \ie, $\cos^2 \theta = 1$,
cannot describe the atmospheric neutrino data
\cite{Fogli:1999qt,Lipari:1999vh}. Figure~\ref{DecayOsc} shows the
survival probability $P_{\mu \mu}$ as a function of the decay rate
$\alpha$ and the sensitivity $L/E$. One can
easily see that the transition from pure oscillation to pure decay is,
in principle, determined by the curve $\frac{\alpha L}{E} = \mathcal{O}(1)$.

\section{Visible decay}
\label{sec:vis}

In this section, we will assume that the neutrino decay products are, in
principle, observable. We know that we may either
detect them separately from the undecayed neutrinos by
different signatures or energies, or indistinguishably to the undecayed
particles with the same signature in the same energy bin\footnote{This strongly
depends on the detector properties. For example, for solar neutrino
decay into antineutrinos with detection in Borexino or
Super-Kamiokande it was discussed in \Ref \cite{Acker:1992eh}.}. We postulated
in \Def~\ref{def:transprob} that these conceptual differences manifest
themselves in a problem-dependent splitting of \eq~(\ref{rdecay}) into
appropriate parts.

\subsection{Transition probability for visible decay}

We use \eqs~(\ref{totapprox1}) and (\ref{papprox}) to first approximation for
$P_{\alpha \beta}^{\rm appearance}$, \ie, the case of maximal one decay or
sufficiently small decay rates. For the calculation of the transition
probability we make the following assumptions: \begin{itemize}
\item Stable decay products or $\alpha_i \ll {E_i \over L}$ for all $i$.
\item The fraction
$\eta_{ij}(L,l,D)$ of decayed particles not escaping detection by a change
of direction does not depend on the decay
position $l$, \ie, $\eta_{ij}(L,l,D)=\eta_{ij}(L,D) \equiv \eta_{ij}$ is a
(real) constant in $l$. In fact, we observed that $\eta_{ij}(L,l,D) \approx 1$
in most cases of Majoron decay (\cf, \Subsec~\ref{ss:vis}).
\item
The masses and decay rates are non-degenerate\footnote{The calculation of
integrals changes for degenerate values of the $\alpha_i$'s and
$m_{i}$'s (see calculation in Appendix \ref{AppearanceTerm}). Nevertheless, we
will calculate limits with degenerate values later on.}.
\end{itemize}

For $\Gamma_{ij}$ and $\Delta_{ij}$ we use the definitions in
\eqs~(\ref{defdelta}) and (\ref{defgamma}).
Applying \eqs~(\ref{totapprox1}) and (\ref{papprox}) and assuming that we
cannot distinguish decay products from undecayed particles, \ie, we add
$P_{\alpha \beta}^0$ and $P_{\alpha \beta}^1$, we have for $P_{\alpha \beta}$
\begin{eqnarray}
P_{\alpha \beta} & \simeq & P_{\alpha \beta}^0 + P_{\alpha
\beta}^1 = P_{\alpha \beta}^{\rm invisible} + P_{\alpha \beta}^{\rm
appearance} \nonumber \\
& = & P_{\alpha \beta}^0 + \int\limits_{l=0}^{L} \left| \left<
\nu_{\beta} | \mathcal{E}(L-l) \mathcal{D}_{-}(L-l) \mathcal{D}_{+}(l,L)
\mathcal{E}(l) \mathcal{D}_{-}(l) | \nu_{\alpha} \right> 
\right|^2 dl.
\label{SumApprox}
\end{eqnarray}

\paragraph*{The invisible decay term.}
$P_{\alpha \beta}^{\rm invisible}$ is given by \eq~(\ref{transprobdecay}),
\ie,
\begin{eqnarray}
P_{\alpha \beta}^{\rm invisible} & = & \sum\limits_i \sum\limits_j \Re
J_{ij}^{\alpha \beta} e^{- \Gamma_{ij}} \nonumber\\
&-& 4 \; \underset{i>j}{\sum\limits_i
\sum\limits_j} \Re J_{ij}^{\alpha \beta} \sin^2 \Delta_{ij} e^{-\Gamma_{ij}}
- 2 \; \underset{i>j}{\sum\limits_i
\sum\limits_j} \Im J_{ij}^{\alpha \beta} \sin 2 \Delta_{ij}
e^{- \Gamma_{ij}}.
\label{Vis1}
\end{eqnarray}

\paragraph*{The appearance term.}
$P_{\alpha \beta}^{\rm appearance}$ evaluates after some algebra (\cf,
Appendix \ref{AppearanceTerm}) to
\begin{eqnarray}
P_{\alpha \beta}^{\rm appearance} & = & \underset{i \neq
j}{\sum\limits_i \sum\limits_j} \underset{k \neq l}{\sum\limits_k
\sum\limits_l} \sqrt{\eta_{ij} \eta_{kl}} {L \over E} \frac{\sqrt{\alpha_{ij}
\alpha_{kl}}}{\left( \Gamma_{jl}-\Gamma_{ik} \right)^2 + 4 \left(
\Delta_{ij} + \Delta_{lk} \right)^2 } \nonumber \\
&\times& \bigg\{ \Re(K_{ijkl}^{\alpha \beta}) \left[ \left( \Gamma_{jl} -
\Gamma_{ik} \right) \left( e^{- \Gamma_{ik}} \cos ( 2 \Delta_{ki} ) -
e^{-\Gamma_{jl}} \cos ( 2 \Delta_{lj} ) \right) \right. \nonumber \\
&-& \left. 2 \left( \Delta_{ij} + \Delta_{lk} \right) \left( e^{-
\Gamma_{ik}} \sin ( 2 \Delta_{ki} ) - e^{-\Gamma_{jl}} \sin ( 2
\Delta_{lj} ) \right) \right] \nonumber \\
&-& \Im(K_{ijkl}^{\alpha \beta}) \left[ \left( \Gamma_{jl} -
\Gamma_{ik} \right) \left( e^{- \Gamma_{ik}} \sin ( 2 \Delta_{ki} ) -
e^{-\Gamma_{jl}} \sin ( 2 \Delta_{lj} ) \right) \right. \nonumber \\
&+& \left. 2 \left( \Delta_{ij} + \Delta_{lk} \right) \left( e^{-
\Gamma_{ik}} \cos ( 2 \Delta_{ki} ) - e^{-\Gamma_{jl}} \cos ( 2 \Delta_{lj} )
\right) \right] \bigg\},
\label{vis2}
\end{eqnarray}
where $K^{\alpha \beta}_{ijkl} \equiv U^*_{\alpha i} U_{\beta j} U_{\alpha k}
U_{\beta l}^*$ is a generalization of $J^{\alpha \beta}_{ij}$.

\subsection{Limiting cases}
\label{LimitingCases}

Let us now consider some special cases for the
general expression for the appearance term in \eq~(\ref{vis2}).

\paragraph*{No oscillations.}

If we ignore neutrino oscillations, \ie, for degenerate masses,
destruction of coherence in decay, or extremely small $\Delta m^2$'s, we
will obtain in the limit $\Delta m^2 \rightarrow 0$
\begin{equation}
P_{\alpha \beta}^{\rm appearance} \rightarrow \underset{i \neq
j}{\sum\limits_i \sum\limits_j} \underset{k \neq l}{\sum\limits_k
\sum\limits_l} \sqrt{\eta_{ij} \eta_{kl}} \frac{2
\sqrt{\alpha_{ij} \alpha_{kl}}}{ \alpha_j +\alpha_l -\alpha_i -\alpha_k}
\Re(K_{ijkl}^{\alpha \beta}) \left( e^{- \Gamma_{ik}} - e^{-\Gamma_{jl}}
\right). \label{appdecay} \end{equation}

\paragraph*{No interference.}

For random phases within superpositions of mass eigenstates, \ie, 
\begin{equation}
 \left| \nu_{\alpha} \right> = a \left| \nu_1 \right> + b e^{i \xi_1} \left|
\nu_2 \right> + c e^{i \xi_2} \left| \nu_3 \right>
\end{equation}
with $a$, $b$, and $c$ some coefficients and $\xi_1$ and $\xi_2$
random phases, the interference terms in \eq~(\ref{vis2}) vanish. We can
include this by introducing $\delta_{ik} \delta_{jl}$ in the sum of
\eq~(\ref{vis2}) and we finally obtain \begin{equation}
P_{\alpha \beta}^{\rm appearance} \rightarrow \underset{i \neq
j}{\sum\limits_i \sum\limits_j} \eta_{ij} \frac{\alpha_{ij}}{
\alpha_j -\alpha_i } \Re(K_{ijij}^{\alpha \beta}) \left( e^{- {\alpha_i L
\over E}} - e^{-{\alpha_j L \over E}} \right).
 \label{appnoint}
\end{equation}

\paragraph*{Only one possible decay channel.}

Assume that there is only one possible decay channel, \ie, $\alpha_{ij}
\equiv \alpha_i \equiv \alpha \neq 0$ and all other $\alpha_{kl} =0$. Then,
we obtain the simplified formula
\begin{equation}
P_{\alpha \beta}^{\rm appearance} \rightarrow \eta_{ij} \Re(K_{ijij}^{\alpha
\beta}) \left( 1 - e^{- {\alpha L \over E}} \right).
 \label{nointdecay}
\end{equation}

\begin{figure}
\begin{center}
\includegraphics*[height=10cm,angle=270]{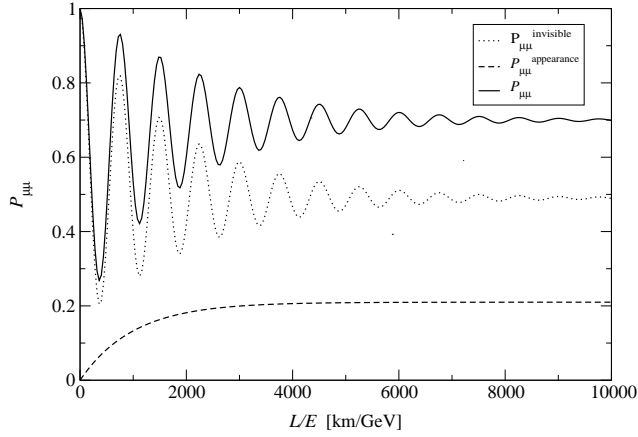}
\end{center}
\caption{\label{VisDecay} The different parts of the survival
probability $P_{\mu \mu}$ plotted as functions of the sensitivity
$L/E$ for the parameter values
$c_{23}^2=0.3$, $\alpha = 0.001 \rm{GeV \over 
km}$, $\eta_{23}=1$, and $\Delta m_{32}^2 = 3.3 \cdot 10^{-3}
\rm{eV}^2$ \cite{Barger:1999bg}.}
\end{figure}

\subsection{Example: Application to atmospheric neutrinos}
\label{atmneutrinos}

Here we will extend the example from \Subsec~\ref{ex:appat1}, where
\eq~(\ref{atmosc}) gives the oscillation formula for decay of
$\nu_{2}$ into an invisible decay product such as a sterile neutrino. Let us
now assume that $\nu_{2}$ decays into the active neutrino $\nu_{3}$.
According to \eq(\ref{SumApprox}), we are now looking for $P_{\mu \mu}^{\rm
appearance}$ in addition to $P_{\mu \mu}^{\rm invisible}$, which can be
calculated from \eq~(\ref{nointdecay}) in this limit. If we assume that both
undecayed and decayed neutrinos are detected indistinguishably, we will obtain
\begin{equation}
P_{\mu \mu} = P_{\mu \mu}^{\rm invisible} + P_{\mu \mu}^{\rm
appearance} \end{equation}
with $P_{\mu \mu}^{\rm invisible}$ in \eq~(\ref{atmosc}) and
\begin{equation}
 \label{atmincoh}
P_{\mu \mu}^{\rm appearance} = \eta_{23} s_{23}^2 c_{23}^2 \left( 1- e^{-
\frac{\alpha L}{E}} \right).
\end{equation}
Here $\alpha$ is the decay rate of $\nu_{2} \rightarrow \nu_{3}$ decay and
$E$ is the energy of the undecayed neutrino. Figure~\ref{VisDecay}
shows the different parts of the survival probability for
the parameter values given in the figure caption. 
One can easily see the effect of the increase of the survival
probability for ${L \over E} \gg {1 \over \alpha}$.

\section{An example for neutrino oscillations of decay products}
\label{sec:nodp}

In this subsection, we will construct a minimal example for observing
neutrino oscillations among decay products.

\subsection{The scenario}

Since we want to create a superposition of {\em out}
states in decay, we have to assume that at least two of the
$g_{ij}$'s, \ie, $\alpha_{ij}$'s, in the interaction
Lagrangian are non-zero. We define a mass hierarchy
$m_3>m_2>m_1$ and only allow the decay chain $\nu_3 \rightarrow \nu_2
\rightarrow \nu_1$, \ie, $\alpha_{32} \equiv \alpha_3 \neq 0$,
$\alpha_{21} \equiv \alpha_2 \neq 0$, and all the other $\alpha_{ij}$'s are
equal to zero.
In order to demonstrate the physical effects we choose suitable parameters.
Hence, we define the $\Delta m^2$'s quite close to each other, \ie, $\Delta
m_{21}^2 = 5 \cdot 10^{-4} \ \rm{eV}^2$, $\Delta m_{32}^2 = 3.3 \cdot 10^{-3}
\ \rm{eV}^2$, maximal mixing, \ie, $\left|
\nu_e \right> = {1 \over \sqrt{3}} \left| \nu_1 \right> + {1 \over \sqrt{3}}
\left| \nu_2 \right> +{1 \over \sqrt{3}} \left| \nu_3 \right>$, and the
$\alpha$'s close to each other, \ie, $\alpha_2 = {1 \over 1000}
\rm{GeV \over km}$ and $\alpha_3 = {1 \over 1100} \rm{GeV \over km}$.
Furthermore, we assume that $\eta_{ij} \approx 1$ for all $i$, $j$ and the
CP-phase $\delta=0$.

\subsection{Transition probabilities}

Let us now look at the disappearance probability of electron neutrinos
$P_{ee}$. {}From \eq~(\ref{transprobdecay}) we obtain the invisible decay
disappearance probability as
\begin{eqnarray}
 P_{ee}^{\rm invisible} & = & J^{ee}_{11} + J^{ee}_{22} e^{-\alpha_2
\frac{L}{E}} +J^{ee}_{33} e^{-\alpha_3 \frac{L}{E}} \nonumber \\
 & + & 2 J_{12}^{ee} e^{-\alpha_2 \frac{L}{E}} \left(
 1- 2 \sin^2 \left( {\Delta m_{21}^2 \over 4} \frac{L}{E} \right) \right)
 \nonumber \\ 
 & + & 2 J_{13}^{ee} e^{-\alpha_3 \frac{L}{E}} \left( 1- 2 \sin^2 \left(
{\Delta m_{31}^2 \over 4} \frac{L}{E} \right) \right) \nonumber \\
 & + & 2 J_{23}^{ee} e^{-{\alpha_2 +\alpha_3 \over 2} \frac{L}{E}} \left(
 1- 2 \sin^2 \left( {\Delta m_{32}^2 \over 4} \frac{L}{E} \right) \right).
\end{eqnarray}

The appearance term is obtained from \eq~(\ref{vis2}). Since
in this case $K^{ee}_{ijkl} = U_{ei}^* U_{ej} U_{ek} U_{el}^* = {1 \over 9}$
and $\Im K =0$, the terms in the sum of \eq~(\ref{vis2}) are equal
for simultaneous $i \leftrightarrow k$ and $j \leftrightarrow l$ index
exchanges. In order to make the result physically more transparent, we
recognize that only a limited number of the terms in the sum are
non-vanishing and split the formula into three parts:
\begin{equation}
 P^{\rm appearance}_{ee} = P^{\rm app,1}_{ee} + P^{\rm app,2}_{ee} +
 P^{\rm int}_{ee}.
\end{equation}
Here $P^{\rm app, 1}_{ee}$ refers to the production of new
$\nu_1$'s by decay,
\begin{equation}
 P^{\rm app, 1}_{ee} = K^{ee}_{2121} \left( 1 - e^{- \alpha_2 \frac{L}{E}}
 \right),
\end{equation} 
$P^{\rm app, 2}_{ee}$ to the production of new
$\nu_2$'s by decay,
\begin{equation}
P^{\rm app, 2}_{ee} = K^{ee}_{3232} {\alpha_3 \over \alpha_2 -
\alpha_3 } \left( e^{-\alpha_3 \frac{L}{E}} - e^{- \alpha_2 \frac{L}{E}}
\right),
\end{equation}
and $P^{\rm int}_{ee}$ to an interference term describing
neutrino oscillations among the decay products $\nu_1$ and
$\nu_2$ (here $K^{ee}_{3221}=K^{ee}_{2132}$ is used),
\begin{eqnarray}
P^{\rm int}_{ee} & = & 4 K^{ee}_{3221} \frac{\sqrt{\alpha_3
\alpha_2}}{\alpha_3^2 + \left( \Delta m_{21}^2 - \Delta m_{32}^2
\right)^2} \nonumber \\
&\times& \Bigg\{ \alpha_3 \left[ e^{- {\alpha_2 \over 2} \frac{L}{E}} \cos
\left( {\Delta m_{21}^2 \over 2} \frac{L}{E} \right) - e^{- {\alpha_2
+ \alpha_3 
\over 2} \frac{L}{E}} \cos \left( {\Delta m_{32}^2 \over 2} \frac{L}{E}
\right) \right] \nonumber \\ & + & \left( \Delta m_{21}^2- \Delta m_{32}^2
\right) \left[ e^{-{\alpha_2 \over 2} \frac{L}{E}} \sin \left( {\Delta
m_{21}^2 \over 2} \frac{L}{E} \right) - e^{- {\alpha_2 + \alpha_3 \over 2}
\frac{L}{E}} \sin \left( {\Delta m_{32}^2 \over 2} \frac{L}{E} \right) 
\right] \Bigg\}. \nonumber\\
\end{eqnarray}

\begin{figure}
\begin{center}
\includegraphics*[height=10cm,angle=270]{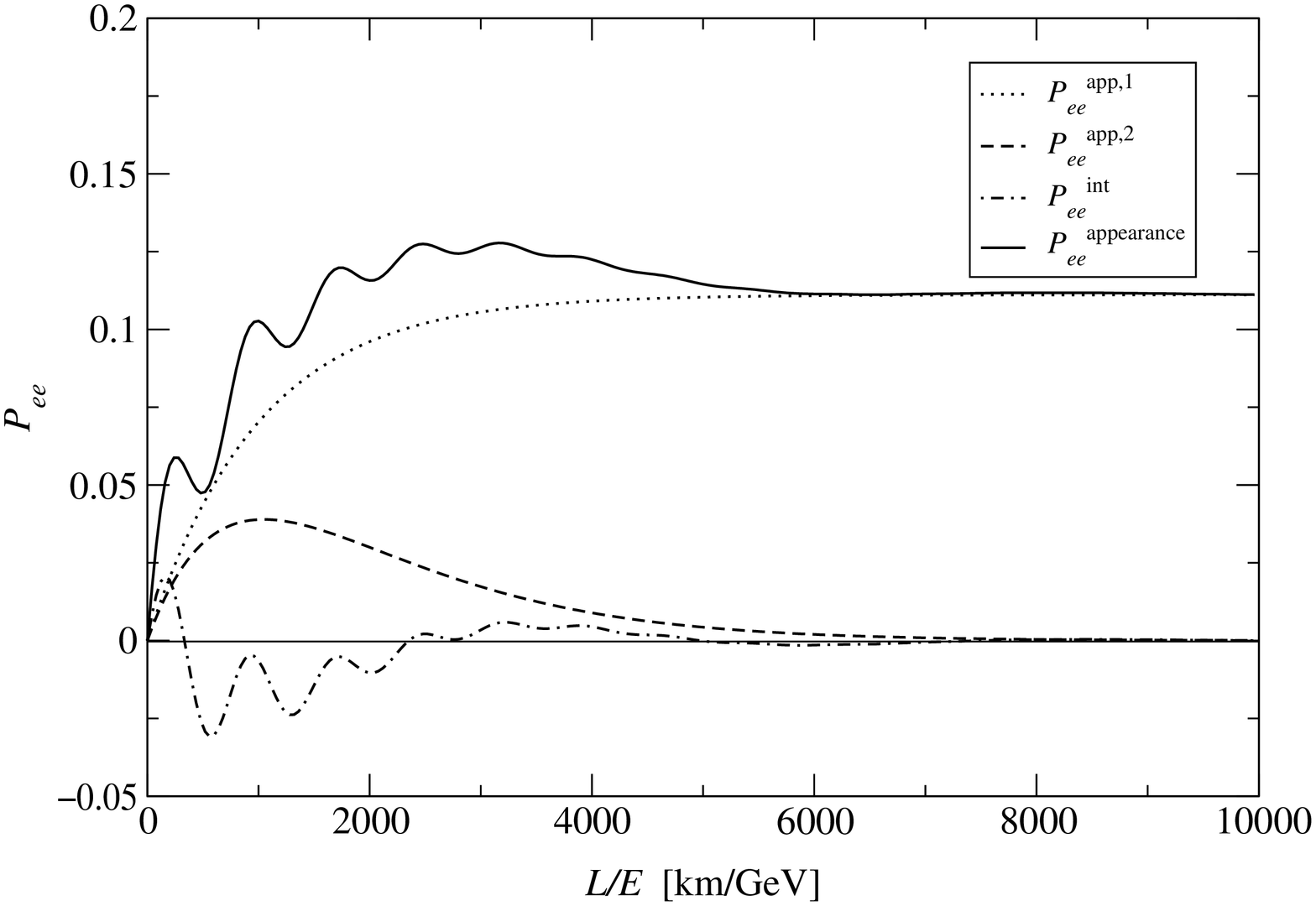}
\end{center}
\caption{\label{TwoSingle} The different parts of the probability $P_{ee}^{\rm
appearance}$ plotted as functions of $L/E$ for the
scenario data.} \end{figure}

\begin{figure}
\begin{center}
\includegraphics*[height=10cm,angle=270]{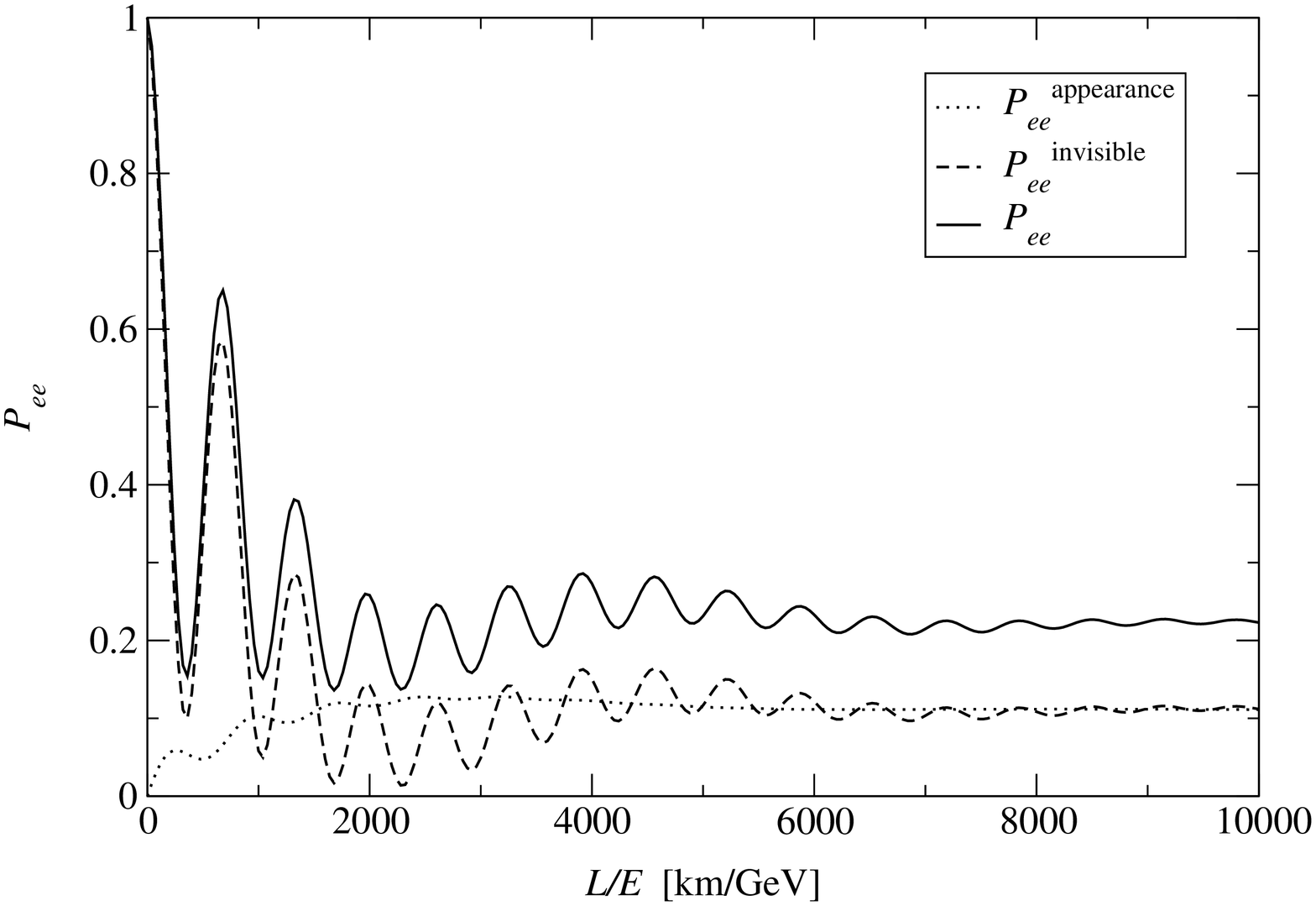}
\end{center}
\caption{\label{TwoTot} The different parts of the probability $P_{ee}$
plotted as functions of $L/E$ for the scenario data.}
\end{figure}

\begin{figure}
\begin{center}
\includegraphics*[height=10cm]{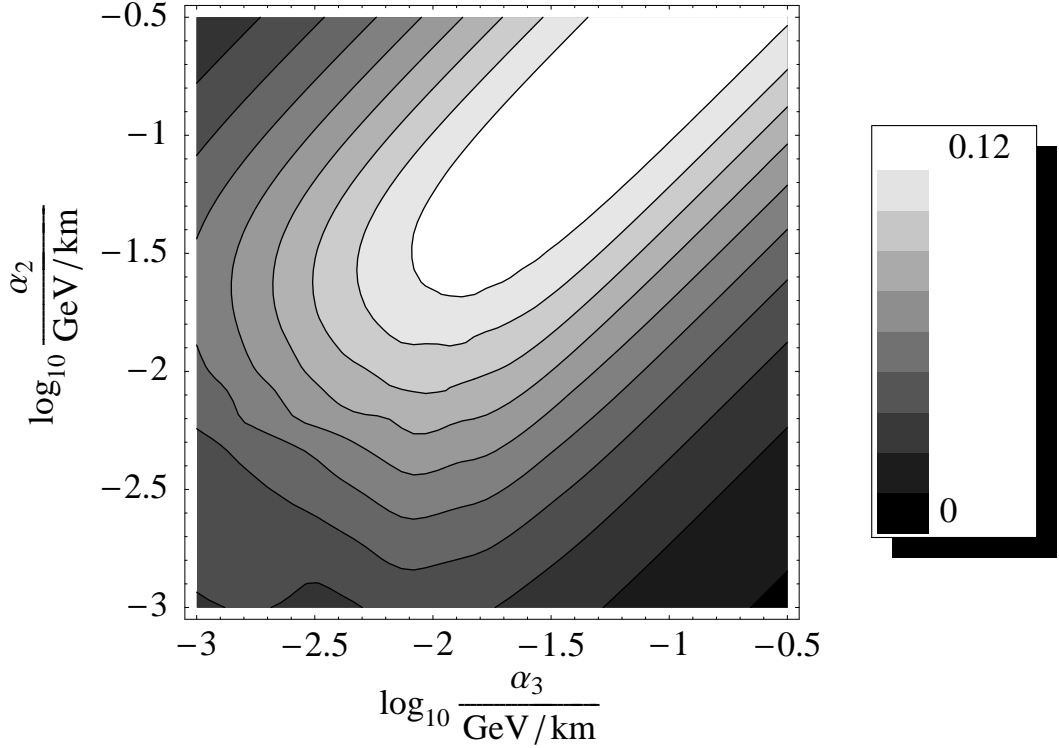}
\end{center}
\caption{\label{TwoMaxInt} Numerical evaluation of the absolute
maximum of $| P_{ee}^{\rm int} |$ for different
decay rates $\alpha_3$ and $\alpha_2$. }
\end{figure}

\subsection{Results and analysis}

Figure~\ref{TwoSingle} shows the different parts of the probability
$P_{ee}^{\rm appearance}$ for the scenario data. The appearance of
$\nu_2$ as decay product is described by $P_{ee}^{\rm app, 2}$. It
grows in the beginning, because it is dominated by $\nu_3 \rightarrow
\nu_2$ decay. Since $\alpha_2>\alpha_3$, the $\nu_2$ is decaying faster
than produced. Therefore, $P_{ee}^{\rm app, 2}$ falls again for
large ${L \over E}$.
The appearance of $\nu_1$ is determined by $P_{ee}^{\rm app,1}$. Since
$\nu_1$ is defined to be stable, $P_{ee}^{\rm app,1}$ grows until it
reaches its asymptotic value $\lim_{x \to \infty} P_{ee}^{\rm app,1} =
K^{ee}_{2121} = {1 \over 9}$.

Describing neutrino oscillations of decay products, \ie, $\nu_1$ and $\nu_2$,
the interference term $P_{ee}^{\rm int}$ basically oscillates as a beat with
the two frequencies determined by $\Delta m_{32}^2 \pm \Delta m_{21}^2$. The
$\Delta m_{32}^2$-dependency comes from the phase propagation before
decay and the $\Delta m_{21}^2$-dependency from the phase propagation
after decay.

Figure~\ref{TwoTot} shows the different
parts of the total survival probability $P_{ee}$. The invisible decay
probability (disappearance) is the ordinary oscillation probability damped by
decay. Note that $P_{ee}^{\rm appearance}$ and $P_{ee}^{\rm invisible}$ can be
only sensibly added if the decay products cannot be distinguished from the
original ones in the detection process. In this case, the detection rates will
be enhanced by a fixed amount at large $L/E$. Neutrino
oscillations of undecayed particles will also vanish in this region. For small
$L/E$ there is a correction due to $\nu_2$ abundance by decay and in addition
due to the interference term.

Neutrino oscillations of decay products, induced by
the oscillations of the interference term, can become unobservable due to
suppression by the mixing angles or the decay rates different by some orders
of magnitude. The first point is quite obvious, since the {\em
interference term} is in general proportional to 
\begin{equation}
 K^{\alpha \beta}_{ijkl} = U^*_{\alpha i} U_{\beta j} U_{\alpha k}
 U^*_{\beta l}, \quad \mbox{where} \; i \neq j, \ k \neq l, \ (i,j) \neq (k,l).
\end{equation} 
This requires that there are coefficients of at least three different
states involved as a product, \ie, {\em all} of them need to be quite large. 
The second point is physically obvious, but harder to see in our
equations. Since the {\em interference} describing neutrino
oscillations of decay products can only be observed if two states are
produced simultaneously in decay, the couplings to both of them need to
be quite large. Large couplings imply large decay rates for both
transitions. If one of them is too large, the corresponding state will
have decayed before oscillations can be observed. Hence, the decay rates
need to be large enough and of the same order of magnitude
in order to obtain reasonable results. A measure for the magnitude of the
interference effects is the absolute maximum of $|P^{\rm int}_{ee} |$. In
\fig~\ref{TwoMaxInt}, we show the result of a numerical evaluation of this
function for different values of the decay rates $\alpha_3$ and $\alpha_2$.
One can easily 
verify that for too small or too different decay rates, this function
gives small values, \ie, we do not see any interference effects. 

\begin{figure}
\begin{center}
\includegraphics*[height=10cm,angle=270]{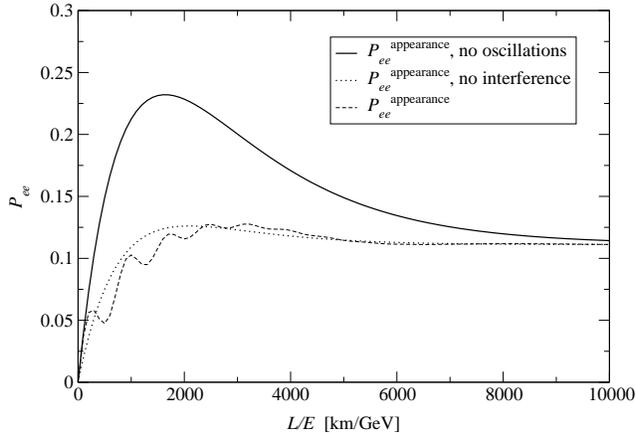}
\end{center}
\caption{\label{TwoApprox} The probability $P_{ee}^{\rm appearance}$ for
different limiting cases plotted as functions of
$L/E$ for the model data.}
\end{figure}

\subsection{Application to limiting cases}

In \Subsec~\ref{LimitingCases}, we studied the
appearance probability in certain limits. Let us now apply these
to our scenario to see what happens if certain terms are assumed to be
negligible.

\paragraph*{No oscillations.}

In this limit, all $\Delta m^2$'s are equal to zero (\cf,
\eq~(\ref{appdecay})). Here we can simply use
\begin{equation}
 P_{ee}^{\rm appearance} = P_{ee}^{\rm app,1} + P_{ee}^{\rm app,2} +
 P_{ee}^{\rm int} \Bigg|_{\Delta m_{ij}^2 \rightarrow 0}.
\end{equation}

\paragraph*{No interference.}

This limit is described by \eq~(\ref{appnoint}), which is equivalent
to neglecting the interference term. We obtain
\begin{equation}
 P_{ee}^{\rm appearance} = P_{ee}^{\rm app,1} + P_{ee}^{\rm app,2}.
\end{equation}
Figure~\ref{TwoApprox} shows the different limits of the appearance
term compared to the original one. In all cases,
we obtain similar values for small and large $L/E$, since for $L/E \ll
(\Delta m^2)^{-1}$ the oscillation terms are ineffective and for $L/E \gg
\alpha^{-1}$ they are suppressed by decay. In the limit of no oscillations,
there is no oscillating behavior of the curve. Nevertheless, there are
interference terms coming from the coherent coupling of the Lagrangian to
different mass eigenstates. Hence, in this case we obtain an enhancement in
the appearance probability. In the limit of no interference, we obtain a
curve similar to the original one, but without oscillations, which are
described by the interference term.

\begin{figure}
\begin{center}
\includegraphics*{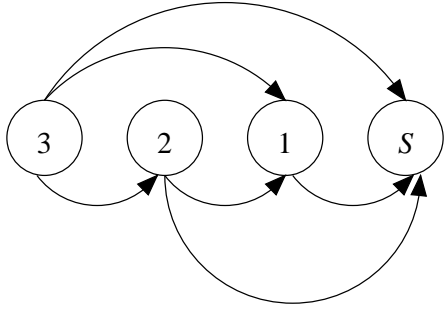}
\end{center}
\caption{\label{DecayModel} Possible decay scenario for four neutrinos.}
\end{figure}

\section{Summary and conclusions}
\label{sec:sc}

In this paper, we have introduced a combined treatment of neutrino decay and
neutrino oscillations in a formal model-independent operator framework.
We have shown that it can be easily applied to special cases such as
atmospheric neutrinos. We have also seen that for Majoron decay active
neutrinos as decay products have to be taken into account in the detection
rates and may even be indistinguishable from the original undecayed particles
by the detector. In addition, in Majoron-like decay models neutrino decay
products may oscillate.

We conclude that the visibility of decay products in Majoron decay models may
crucially affect the detection rates depending on the properties of the
detector. For other decay models only a certain fraction
of the decay products may arrive at the detector due to kinematics.
Moreover, neutrino oscillations of decay products are observable for
relatively large neutrino flavor mixing only.

One can show that for the combination of neutrino decay and neutrino
oscillations, the dimension of the parameter space is equal to ${3
\over 2} n (n-1)$, where ${1 \over 2} n (n-1)$ parameters come from
decay and $n (n-1)$ from oscillations. Here $n$
denotes the number of neutrino flavors. Hence, the number of
parameters is increased when neutrino decay is included. Thus, especially in
multi-neutrino scenarios, effects not consistent with the conventional
neutrino oscillation framework could be described by a combined neutrino decay
and neutrino oscillation approach.

As a realistic scenario, we could, for example, use a four-neutrino
framework with a decoupled sterile neutrino $\nu_S$ (\cf,
\eq~(\ref{decoupled}))
and a mass hierarchy $m_3>m_2>m_1>m_S$ (\cf, \fig~\ref{DecayModel}). In such a
scenario, we have up to six decay rates in addition to the usual
mixing parameters. {}From other conditions such as the observability
of supernova 
neutrinos, we then would need to constrain the lifetimes for individual
$\nu_i$'s (\eg, $\nu_1$ almost stable). Since the maximal number of possible
decays is $N_{\max}=3$ in this scenario, the exact transition probability
$P_{\alpha \beta}$ without lifetime constraints would be determined by
$P_{\alpha \beta}^0$, $P_{\alpha \beta}^1$, $P_{\alpha \beta}^2$, and
$P_{\alpha \beta}^3$ in \eq~(\ref{rdecay}).

Since non-zero neutrino masses imply both neutrino decay and
neutrino oscillations, neutrino decay may become interesting not only as an
alternative to neutrino oscillations but also as a part of the whole
neutrino scenario. Even small $g_{ij}$'s may affect detection rates of long
traveling neutrinos such as supernova or primordial ones. Flux
measurements (\eg, of supernova neutrinos) cannot always be used to
extract neutrino mass limits reliably, because of different path
lengths of undecayed and decayed neutrinos. Thus, dispersion may
be mimiced by decay.

As a final conclusion, our operator formalism is a
generalization for a combined treatment of neutrino decay and neutrino
oscillations could be of interest to those, who are interested in analyzing
neutrino data in a more general way. In addition, it may also be
applied in other oscillation scenarios outside of neutrino physics,
where coherence is not destroyed in decay.

\ack
We would like to thank Evgeny Akhmedov for useful discussions and J{\"o}rn
Kersten for proof-reading the manuscript.

This work was supported by the Swedish Foundation for International
Cooperation in Research and Higher Education (STINT), the Wenner-Gren
Foundations, the ``Studienstiftung des deutschen Volkes'' (German National
Merit Foundation), and the ``Sonderforschungsbereich 375 f{\"u}r
Astro-Teilchenphysik der Deutschen Forschungsgemeinschaft''.

\begin{appendix}

\section{Calculation of the appearance term}
\label{AppearanceTerm}

We are using \eq~(\ref{papprox}) as an approximation for the
appearance term:
\begin{eqnarray}
P_{\alpha \beta}^{\rm appearance} & \simeq & \int\limits_{l=0}^{L} \left|
\underbrace{\left< \nu_{\beta} | \mathcal{E}(L-l) \mathcal{D}_{-}(L-l)
\mathcal{D}_{+}(l,L) \mathcal{E}(l) \mathcal{D}_{-}(l) | \nu_{\alpha} \right>
}_{A_{\alpha \beta}} \right|^2 dl = \nonumber \\
& = & \int\limits_{l=0}^{L} \left| A_{\alpha \beta} \right|^2 dl,
\end{eqnarray}
where $A_{\alpha \beta}$ is the (differential) transition amplitude.
Applying \eqs~(\ref{DMinus}), (\ref{DPlus}), (\ref{Evol}), and $\left|
\nu_{\alpha} \right> = \sum_i U_{\alpha i}^* \left| \nu_i \right>$, we obtain
for the transition amplitude\footnote{Here we already neglect the
random phase factor $e^{i \xi}$ from the operator $\mathcal{D}_{+}$, since it
will cancel anyway by taking its absolute value squared in the transition
probability.}
\begin{equation}
A_{\alpha \beta} = \underset{i \neq j}{\sum\limits_{i}
\sum\limits_{j}} U_{\alpha i}^*
U_{\beta j} \ e^{-i E_j (L-l)} \ e^{- \frac{\alpha_j (L-l)}{2 E_j}}
\ \sqrt{\alpha_{ij} \over E_i} \sqrt{\eta_{ij}} \ e^{-i E_i l} \ e^{-
\frac{\alpha_i l}{2 E_i}}.
\end{equation}
In this equation, we already have evaluated the creation and annihilation
operators for readability.
Now we use the approximation for relativistic neutrinos $E_i
\simeq p + {m_i^2 \over 2 p} \simeq p + {m_i^2 \over 2 E}$. For neutrino
oscillations the three-momentum $p$ only gives a common phase factor, which
will cancel. For neutrino decay, the exponentials $e^{ - \alpha_i l
\over 2 E_i}$ and $e^{ - \alpha_j (L-l) \over 2 E_j}$
are real, and thus $E_i \simeq p \simeq E$ in the lowest order nontrivial
approximation. Using the relativistic approximations as well as the
definition $\Delta m_{ab}^2 \equiv m_a^2 - m_b^2$ then yields
\begin{equation}
A_{\alpha \beta} = \underset{i \neq j}{\sum\limits_{i}
\sum\limits_{j}} \ \underbrace{U_{\alpha i}^* 
U_{\beta j} \sqrt{\eta_{ij}} \sqrt{\alpha_{ij} \over E} e^{-
\frac{\alpha_j L}{2 E}} e^{-i \frac{m_j^2}{2E} L} }_{l{\rm -independent}} \
\underbrace{ e^{ \frac{(\alpha_j - \alpha_i)
l}{2 E} }e^{i \frac{\Delta m_{ji}^2}{2E} l} }_{l{\rm -dependent}}.
\end{equation}
For the appearance probability we obtain
\begin{eqnarray}
P_{\alpha \beta}^{\rm appearance} & = & \int\limits_{l=0}^{L}
\left| A_{\alpha \beta} \right|^2 dl = \int\limits_{l=0}^{L}
A_{\alpha \beta} A_{\alpha \beta}^* dl
 = \underset{i \neq j}{\sum\limits_{i} \sum\limits_{j}} \underset{k
\neq l}{\sum\limits_{k} \sum\limits_{l}} \ \underbrace{U_{\alpha
i}^* U_{\beta j} U_{\alpha k} 
U_{\beta l}^*}_{\equiv K_{ijkl}^{\alpha \beta}} \nonumber\\
&\times& \sqrt{\eta_{ij} \eta_{kl}}
\sqrt{\alpha_{ij} \over E} \sqrt{\alpha_{kl} \over E} e^{-
\frac{\alpha_j +\alpha_l}{2 
E} L} e^{-i \frac{\Delta m_{jl}^2}{2E} L}
\int\limits_{l=0}^{L}
e^{ \frac{\alpha_j +\alpha_l - \alpha_i -\alpha_k}{2 E} l} e^{i
\frac{\Delta m_{ji}^2 - \Delta m_{lk}^2 }{2E} l} dl, \nonumber\\
\label{appp1}
\end{eqnarray}
where the $\eta_{ij}$'s are assumed to be real.
For further evaluation we use
the integral ($a$, $b$ real)
\begin{equation}
\int\limits_0^L e^{(a+bi) l} dl = {1 \over a+bi} \left( e^{(a+bi) L}
-1 \right) = \frac{a-bi}{a^2+b^2} \left( e^{(a+bi) L} -1 \right)
\label{lint}
\end{equation}
for $a+bi \neq 0$. Comparison with \eq~(\ref{appp1}) shows that the latter
condition implies
\begin{equation}
\alpha_j + \alpha_l - \alpha_i - \alpha_k + i \left(
m_j^2-m_i^2-m_l^2+m_k^2 \right) \neq 0.
\end{equation}
Since $i \neq j \wedge k \neq l$ by the summation rules, this can only
happen for $(i=k) \wedge (j=l) \wedge (\alpha_i = \alpha_j)$ for
non-degenerate $m$'s. Here we assume the $\alpha$'s to be non-degenerate
for simplicity, so that we can apply \eq~(\ref{lint}).
Using the same abbreviations as for invisible decay in
\eqs~(\ref{defdelta}) and (\ref{defgamma}), we can continue evaluating
\eq~(\ref{appp1}) with \eq~(\ref{lint}) by identifying $a= \frac{\alpha_j +
\alpha_l - \alpha_i - \alpha_k}{2E} = \Gamma_{jl} - \Gamma_{ik}$ and $b =
\frac{\Delta m_{ji}^2 - \Delta m_{lk}^2}{2E} = 2 \left( \Delta_{ji}+
\Delta_{kl} \right)$
\begin{eqnarray}
P_{\alpha \beta}^{\rm
appearance} & = & \underset{i \neq j}{\sum\limits_{i} \sum\limits_{j}}
\underset{k \neq l}{\sum\limits_{k} \sum\limits_{l}} K_{ijkl}^{\alpha
\beta} \sqrt{\eta_{ij} \eta_{kl}} \sqrt{\alpha_{ij} \alpha_{kl}} {L
\over E} e^{- \frac{\alpha_j +\alpha_l}{2 E} L} e^{-i \frac{\Delta
m_{jl}^2}{2E} L}
\nonumber \\
&\times& \frac{\Gamma_{jl} - \Gamma_{ik} - 2 i \left( 
\Delta_{ji} + \Delta_{kl} \right)}{\left(\Gamma_{jl} - \Gamma_{ik}\right)^2 +4
\left( \Delta_{ji} + \Delta_{kl} \right)^2} \left( e^{ \left(
\alpha_j + \alpha_l - \alpha_i - \alpha_k + i \left( \Delta m_{ji}^2-
\Delta m_{lk}^2 \right) \right) {L \over 2E}} -1 \right) \nonumber \\
& = & \underset{i \neq j}{\sum\limits_{i} \sum\limits_{j}}
\underset{k \neq l}{\sum\limits_{k} \sum\limits_{l}}
K_{ijkl}^{\alpha \beta} \sqrt{\eta_{ij} \eta_{kl}} 
\sqrt{\alpha_{ij} \alpha_{kl}} {L \over E} \frac{\Gamma_{jl} -
\Gamma_{ik} - 2 i \left( \Delta_{ji} + \Delta_{kl}
\right)}{\left(\Gamma_{jl} - \Gamma_{ik}\right)^2 +4 \left(
\Delta_{ji} + \Delta_{kl} \right)^2} \nonumber \\
&\times& \left( e^{-\Gamma_{ik}} e^{2 i \Delta_{ki}} -
e^{-\Gamma_{jl}} e^{2 i \Delta_{lj}} \right).
\label{appp2}
\end{eqnarray}
Because $P_{\alpha \beta}^{\rm appearance}$ is real, we have $P_{\alpha
\beta}^{\rm appearance} = \Re P_{\alpha \beta}^{\rm appearance}$. For $\phi$
real, $y$ and $z$ complex, it can be shown that
\begin{eqnarray}
\Re (z e^{i \phi}) & = & \Re (z) \cos \phi - \Im (z) \sin \phi, \\
\Re (i y e^{i \phi}) & = & - \Im (y) \cos \phi - \Re (y) \sin \phi.
\end{eqnarray}
We can use this to find
\begin{eqnarray}
 \Re \left( K (c+d i) e^{2 i \Delta} \right) & = & \Re (K) c \cos 2
 \Delta - \Im (K) d \sin 2 \Delta \nonumber \\
 & - & \Im (K) c \cos 2 \Delta - \Re (K) d \sin 2 \Delta
\end{eqnarray}
for $c$, $d$ real. In comparison with \eq~(\ref{appp2}), we identify $c =
\Gamma_{jl} - \Gamma_{ik}$ and $d = 2 ( \Delta_{ij} + \Delta_{lk}
)$. Applying
this formula and re-grouping in $\Re K$ and $\Im K$ yields the final
result in \eq~(\ref{vis2}).

\end{appendix}

\bibliographystyle{h-elsevier}
\bibliography{references}

\end{document}